\documentclass[utf8]{frontiersSCNS} 

\usepackage{url,hyperref,lineno,microtype,subcaption}
\usepackage[onehalfspacing]{setspace}
\usepackage[T1]{fontenc}
\usepackage{bm}
\usepackage{amsmath,etoolbox}
\usepackage{graphicx}                    
\usepackage{color}                       
\usepackage{breakurl}                         
\usepackage{amssymb}
\usepackage{amsbsy}
\usepackage{latexsym}
\usepackage{marvosym}
\usepackage{enumerate}
\usepackage{natbib}
\usepackage{textcomp}
\usepackage{float}
\usepackage{afterpage}
\DeclareMathAlphabet{\mathsfit}{\encodingdefault}{\sfdefault}{m}{sl}
\SetMathAlphabet{\mathsfit}{bold}{\encodingdefault}{\sfdefault}{bx}{sl}
\newcommand{\tens}[1]{\bm{\mathsfit{#1}}}

\def\keyFont{\fontsize{8}{11}\helveticabold }
\def\firstAuthorLast{Zhelyazkov, Chandra and Joshi} 

\def\Address{$^\textit{1}$Faculty of Physics, Sofia University, 1164 Sofia, Bulgaria \\
$^\textit{2}$Department of Physics, DSB Campus, Kumaun University, Nainital 263\,002, India }
\def\corrAuthor{Ivan Zhelyazkov}

\def\corrEmail{izh@phys.uni-sofia.bg}

\begin{document}
\onecolumn
\firstpage{1}

\title[Kelvin--Helmholtz Instability in Rotating Jets]{How Rotating Solar Atmospheric Jets Become Kelvin--Helmholtz Unstable}

\author[\firstAuthorLast ]{Ivan Zhelyazkov$^{\textit{\textbf{1,*}}}$, Ramesh Chandra$^{\textit{\textbf{2}}}$ and Reetika Joshi$^{\textit{\textbf{2}}}$} 
\address{} 
\correspondance{} 

\extraAuth{}

\maketitle

\begin{abstract}
\section{}
$\!\!$Recent observations support the propagation of a number of magnetohydrodynamic (MHD) modes which, under some conditions, can become unstable and the developing instability is the Kelvin--Helmholtz instability (KHI).  In its nonlinear stage the KHI can trigger the occurrence of wave turbulence which is considered as a candidate mechanism for coronal heating.  We review the modeling of tornado-like phenomena in the solar chromosphere and corona as moving weakly twisted and spinning cylindrical flux tubes, showing that the KHI rises at the excitation of high-mode MHD waves.  The instability occurs within a wavenumber range whose width depends on the MHD mode number \emph{m}, the plasma density contrast between the rotating jet and its environment, and also on the twists of the internal magnetic field and the jet velocity.  We have studied KHI in two twisted spinning solar polar coronal hole jets, in a twisted rotating jet emerging from a filament eruption, and in a rotating macrospicule.  The theoretically calculated KHI development times of a few minutes for wavelengths comparable to the half-widths of the jets are in good agreement with the observationally determined growth times only for high order (10 $\mathrm{\leqslant}$ \emph{m} $\mathrm{\leqslant}$ 65) MHD modes. Therefore, we expect that the observed KHI in these cases is due to unstable high-order MHD modes.

\tiny
 \keyFont{ \section{Keywords:} magnetic fields, magnetohydrodynamics (MHD), solar jets, MHD waves and instabilities, numerical methods} 
\end{abstract}

\section{Introduction}
\label{sec:intro}
Solar jets are ubiquitous in the solar atmosphere and recent observations have revealed that they are related to small scale filament eruptions.  They are continuously observed by the Extreme-ultraviolet Imaging Spectrometer (EIS) \citep{Culhane2007} on board \emph{Hinode\/} \citep{Kosugi2007} satellite, Atmospheric Imaging Assembly (AIA) \citep{Lemen2012}, on board the \emph{Solar Dynamics Observatory\/} (\emph{SDO}) \citep{Pesnell2012}, as well as from the \emph{Interface Region Imaging Spectrograph\/} (\emph{IRIS}) \citep{Depontieu2014} alongside the Earth-based solar telescopes.  The physical parameters of various kinds of solar jets have been reported in a series of articles, see for instance, \citet{Schmieder2013}, \citet{Sterling2015}, \citet{Panesar2016a}, \citet{Chandra2017}, \citet{Joshi2017}, and references cited in.  It was established that more of the solar jets possess rotational motion.  Such tornado-like jets, termed macrospicules, were firstly detected in the transition region by \citet{Pike1998} using observations by the \emph{Solar and Heliospheric Observatory\/} {(\emph{SOHO}) \citep{Domingo1995}}.  Rotational motion in macrospicules was also explored by \citet{Kamio2010}, \citet{Curdt2011}, \citet{Bennett2015}, \citet{Kiss2017}, and \citet{Kiss2018}.  Type II spicules, according to \citet{DePontieu2012} and \citet{Martinez-Sykora2013}, along with the coronal hole EUV jets \citep{Nistico2009,Liu2009,Nistico2010,Shen2011,Chen2012,Hong2013,Young2014a,Young2014b,Moore2015}, and X-ray jets \citep{Moore2013}, can rotate, too.  Rotating EUV jet emerging from a swirling flare \citep{Zhang2014} or formed during a confined filament eruption \citep{Filippov2015} confirm once again the circumstance that the rotational motion is a common property of many kinds of jets in the solar atmosphere.

The first scenario for the numerical modeling of hot X-ray jets was reported by \citet{Heyvaerts1977} and the basic idea was that a bipolar magnetic structure emerges into a unipolar pre-existing magnetic field and reconnects to form hot and fast jets that are emitted from the interface between the fields into contact.  Later on, by examining many X-ray jets in \emph{Hinode}/X-Ray Telescope coronal X-ray movies of the polar coronal holes, \citet{Moore2010} found that there is a dichotomy of polar X-ray jets, namely ``standard'' and ``blowout'' jets exist.  \citet{Fang2014} studied the formation of rotating coronal jets through numerical simulation of the emergence of a twisted magnetic flux rope into a pre-existing open magnetic field.  Another scenario for the nature of solar jets was suggested by \citet{Sterling2015}, according to which the X-ray jets are due to flux cancellation and/or ``mini-eruptions'' rather than emergence. An alternative model for solar polar jets due to an explosive release of energy via reconnection was reported by \citet{Pariat2009}.  Using three-dimensional MHD simulations, the authors demonstrated that this mechanism does produce massive, high-speed jets.  In subsequent two articles \citep{Pariat2015,Pariat2016}, Pariat and co-authors, presented several parametric studies of a three-dimensional numerical MHD model for straight and helical solar jets.  On the other side, \citet{Panesar2016b} have shown that the magnetic flux cancellation can trigger the solar quiet-region coronal jets and they claim that the coronal jets are driven by the eruption of a small-scale filament, called a ``minifilament.''  The small-scale chromospheric jets, like microspicules, were first numerically modeled by \citet{Murawski2011}.  Using the FLASH code, they solved the two-dimensional ideal MHD equations to model a macrospicule, whose physical parameters match those of a solar spicule observed.  Another mechanism for the origin of macrospicules was proposed by \citet{Kayshap2013}, who numerically modeled the triggering of a macrospicule and a jet.

It is natural to expect, that solar jets, being magnetically structured entities, should support the propagation of different type of MHD waves: fast and slow magnetoacoustic waves and torsional Alfv\'en waves.  All these waves are usually considered as normal MHD modes traveling along the jet.  Owing to the presence of a velocity shear near the jet--surrounding plasma interface, every jet can become unstable and the most universal instability which emerges is the Kelvin--Helmholtz (KH) one.  The simplest configuration at which one can observe the KHI is the two semi-infinite incompressible magnetized plasmas flowing with different velocities provided that the thin velocity shear at the interface exceeds some critical value \citep{Chandrasekhar1961}.  Recently, \citet{Cheremnykh2018a} theoretically established that shear plasma flows at the boundary of plasma media can generate eight MHD modes, of which only one can be unstable due to the development of the KHI.  \citet{Ismayilli2018} investigated a shear instability of the KH type in a plasma with temperature anisotropy under the MHD approximation.  The KHI of the magnetoacousic waves propagating in a steady asymmetric slab, and more specifically the effect of varying density ratios was explored by \citet{Barbulescu2018}.  A very good review on the KHI in the solar atmosphere, solar wind, and geomagnetosphere in the framework of ideal MHD the reader can find in \citealp{Mishin2016}.

In cylindrical geometry, being typical for the solar jets, the KHI exhibits itself as a vortex sheet running on the jet--environment boundary, which like in the flat geometry, is growing in time if the axial velocity of the jet in a frame of reference attached to the surrounding plasma exceeds a threshold value \citep{Ryu2000}.  In its nonlinear stage, the KHI trigger the wave turbulence which is considered as one of the main heating mechanisms of the solar corona \citep{Cranmer2015}. The development of the KHI in various cylindrical jet--environment configurations has been studied in photospheric jets \citep{Zhelyazkov2012a}, in solar spicules \citep{Zhelyazkov2012,Ajabshirizadeh2015,Ebadi2016}, in high-temperature and cool solar surges \citep{Zhelyazkov2015a,Zhelyazkov2015b}, in magnetic tubes of partially ionized compressible plasma \citep{Soler2012}, in EUV chromospheric jets \citep{Zhelyazkov2016,Bogdanova2018}, in soft X-ray jets \citep{Vasheghani2009,Zhelyazkov2017}, and in the twisted solar wind flows \citep{Zaqarashvili2014}.  A review on KHI in the solar atmosphere, including some earlier studies, the reader can find in \citealp{Zhelyazkov2015}.

The first modeling of the KHI in a rotating cylindrical magnetized plasma jet was done by \citet{Bondenson1987}.  Later on, \citet{Bodo1989,Bodo1996} carried out a study of the stability of flowing cylindrical jet immersed in constant magnetic field $\boldsymbol{B}_0$.  The authors used the standard procedure for exploring the MHD wave propagation in cylindrical flows considering that all the perturbations of the plasma pressure $p$, fluid velocity $\boldsymbol{v}$, and magnetic field $\boldsymbol{B}$, are ${\propto}\exp[\mathrm{i}(-\omega t + kz + m\theta)]$.  Here, $\omega$ is the angular wave frequency, $k$ the propagating wavenumber, and $m$ the azimuthal mode number.  Using the basic equations of ideal magnetohydrodynamics, \citet{Bodo1989,Bodo1996} derived a Bessel equation for the pressure perturbation and an expression for the radial component of the fluid velocity perturbation.  The found solutions in both media (the jet and its environment) are merged at the perturbed tube boundary through the conditions for continuity of the total (thermal plus magnetic) pressure and the Lagrangian displacement.  The latter is defined as the ratio of radial velocity perturbation component and the angular frequency in the corresponding medium.  The obtained dispersion relation is used for examining the stability conditions of both axisymmetric, $m = 0$ \citep{Bodo1989}, and non-axisymmetric, $|m| \geqslant 1$ modes \citep{Bodo1996}.  In a recent article, \citet{Bodo2016} performed a linear stability analysis of magnetized rotating cylindrical jet flows in the approximation of zero thermal pressure.  They focused their analysis on the effect of rotation on the current driven mode and on the unstable modes introduced by rotation.  In particular, they found that rotation has a stabilizing effect on the current driven mode only for rotation velocities of the order on the Alfv\'en speed.  The more general case, when both the magnetic field and jet flow velocity are twisted, was studied by \citet{Zaqarashvili2015} and \citet{Cheremnykh2018b}, whose dispersion equations for modes with $m \geqslant 2$, represented in different ways, yield practically identical results.

The main goal of this review article is to suggest a way of using the wave dispersion relation derived in \citealp{Zaqarashvili2015} to study the possibility for the rising and development of KHI in rotating twisted solar jets.  Among the enormous large number of observational studies of rotating jets with different origin or nature, we chose those which provide the magnitudes of axial and rotational speeds, jet width and height alongside the typical plasma parameters like electron number densities and electron temperature of the spinning structure and its environment.  Thus, the targets of our exploration are: (i) the spinning coronal hole jet of 2010 August 21 \citep{Chen2012}; (ii) the rotating coronal hole jet of 2011 February 8 \citep{Young2014a}, (iii) the twisted rotating jet emerging from a filament eruption on 2013 April 10--11 \citep{Filippov2015}, and (iv) the rotating macrospicule observed by \citet{Pike1998} on 1997 March 8.

The paper is organized as follows: in the next section, we discuss the geometry of the problem, equilibrium magnetic field configuration and basic physical parameters of the explored jets.  Section~3 is devoted to a short, concise, derivation of the wave dispersion relation.  Section~4 deals with numerical results for each of the four jets and contain the available observational data.  In the last Section~5, we summarize the main findings in our research and outlook the further improvement of the used modeling.

\section{The Geometry, Magnetic Field, and Physical Parameters in a Jet Model}
\label{sec:geometry}
We model whichever jet as an axisymmetric cylindrical magnetic flux tube with radius $a$ and electron number density $n_\mathrm{i}$ (or equivalently, homogeneous plasma density $\rho_\mathrm{i}$) moving with velocity $\boldsymbol{U}$.  We consider that the jet environment is a rest plasma with homogeneous density $\rho_\mathrm{e}$ immersed in a homogeneous background magnetic field $\boldsymbol{B}_\mathrm{e}$.  This field, in cylindrical coordinates ($r, \phi, z$), possesses only an axial component, \emph{i.e.},  $\boldsymbol{B}_\mathrm{e} = (0, 0, B_\mathrm{e})$.  (Note that the label `i' is abbreviation for \emph{interior}, and the label `e' denotes \emph{exterior}.)  The magnetic field inside the tube, $\boldsymbol{B}_\mathrm{i}$, and the jet velocity, $\boldsymbol{U}$, we assume, are uniformly twisted and are given by the vectors
\begin{equation}
\label{eq:Bi-U}
    \boldsymbol{B}_\mathrm{i} = \left( 0, B_{\mathrm{i}\phi}(r), B_{\mathrm{i}z} \right) \qquad \mbox{and} \qquad \boldsymbol{U} = \left( 0, U_{\phi}(r), U_{z} \right),
\end{equation}
respectively.  We note, that $B_{\mathrm{i}z}$ and $U_z$, are constant.  Concerning the azimuthal magnetic and flow velocity components, we suppose that they are linear functions of the radial position $r$ and evaluated at $r = a$ they correspondingly are equal to $B_{\mathrm{i}\phi}(a) \equiv B_{\phi} = Aa$ and $U_{\phi} = \Omega a$, where $A$ and $\Omega$ are constants.  Here, $\Omega$ is the jet angular speed, deduced from the observations.  Hence, in equilibrium, the rigidly rotating plasma column, that models the jet, must satisfy the following force-balance equation (see, \emph{e.g.}, \citealp{Chandrasekhar1961}; \citealp{Goossens1992})
\begin{equation}
\label{eq:forceeq}
    \frac{\mathrm{d}}{\mathrm{d}r}\left( p_\mathrm{i} + \frac{B_\mathrm{i}^2}{2\mu} \right) = \frac{\rho_\mathrm{i}U_\phi^2}{r} - \frac{B_{\mathrm{i}\phi}^2}{\mu r},
\end{equation}
where $\mu$ is the plasma permeability and $p_\mathrm{t} = p_\mathrm{i} + B_\mathrm{i}^2/2\mu$ with $B_\mathrm{i}^2 = B_{\mathrm{i}\phi}^2(r) + B_{\mathrm{i}z}^2$ is the total (thermal plus magnetic) pressure.  According to equation (\ref{eq:forceeq}), the radial gradient of the total pressure should balance the centrifugal force and the force owing to the magnetic tension.  After integrating equation~(\ref{eq:forceeq}) from $0$ to $a$, taking into account the linear dependence of $U_\phi$ and $B_{\mathrm{i}\phi}$ on $r$, we obtain that
\[
    p_\mathrm{t}(a) = p_\mathrm{t}(0) + \frac{1}{2}\rho_\mathrm{i}U_\phi^2(a) - \frac{B_{\mathrm{i}\phi}^2(a)}{2\mu},
\]
where $p_\mathrm{t}(0) = p_1(0) + B_{\mathrm{i}z}^2/2\mu$.  (Integrating equation~(\ref{eq:forceeq}) from $0$ to any $r$ one can find the radial profile of $p_\mathrm{t}$ inside the tube.  Such an expression of $p_\mathrm{t}(r)$, obtained, however, from an integration of the momentum equation for the equilibrium variables, have been obtained in \citealp{Zhelyazkov2018a}---see Eq.~(2) there.)  It is clear from a physical point of view that the internal total pressure (evaluated at $r = a$) must be balanced by the total pressure of the surrounding plasma which implies that
\[
    p_1(0) + \frac{B_{\mathrm{i}z}^2}{2\mu} - \frac{B_{\mathrm{i}\phi}^2(a)}{2\mu} + \frac{1}{2}\rho_\mathrm{i}U_\phi^2(a) = p_\mathrm{e} + \frac{B_{\mathrm{e}}^2}{2\mu}.
\]
This equation can be presented in the form
\begin{equation}
\label{eq:pbeq}
    p_1(0) + \frac{1}{2}\rho_\mathrm{i}U_\phi^2(a) + \frac{B_{\mathrm{i}z}^2}{2\mu}\left( 1 - \varepsilon_1^2 \right) = p_\mathrm{e} + \frac{B_\mathrm{e}^2}{2\mu},
\end{equation}
where $p_1(0)$ is the thermal pressure at the magnetic tube axis, and $p_\mathrm{e}$ denotes the thermal pressure in the environment.  In the pressure balance equation~(\ref{eq:pbeq}), the number $\varepsilon_1 \equiv B_{\phi}/B_{\mathrm{i}z} = Aa/B_{\mathrm{i}z}$ represents the magnetic field twist parameter.  Similarly, we define $\varepsilon_2 \equiv U_\phi/U_z$ as a characteristics of the jet velocity twist.  We would like to underline that the choice of plasma and environment parameters must be such that the total pressure balance equation~(\ref{eq:pbeq}) is satisfied.  In our case, the value of $\varepsilon_2$ is fixed by observationally measured rotational and axial velocities while the magnetic field twist, $\varepsilon_1$, has to be specified when using equation~(\ref{eq:pbeq}).  We have to note that equation~(\ref{eq:pbeq}) is a corrected version of the pressure balance equation used in \citealp{Zhelyazkov2018a,Zhelyazkov2018b}.

From measurements of $n$ and $T$ for similar coronal hole EUV jets \citep{Nistico2009,Nistico2010}, we take $n$ inside the jet to be $n_\mathrm{i} = 1.0 \times 10^9$~cm$^{-3}$, and assume that the electron temperature is $T_\mathrm{i} = 1.6$~MK.  The same quantities in the environment are respectively $n_\mathrm{e} = 0.9 \times 10^9$~cm$^{-3}$ and $T_\mathrm{e} = 1.0$~MK.  Note that the electron number density of the blowout jet observed by \citet{Young2014a} is in one order lower.  The same applies for its environment.  We consider that the background magnetic field for both hole coronal jets is $B_\mathrm{e} = 3$~G.  The values of $n$ and $T$ of the rotating jet emerging from a filament eruption, observed by \citet{Filippov2015}, were evaluated by us and they are $n_\mathrm{i} = 4.65 \times 10^9$~cm$^{-3}$ and $T_\mathrm{i} = 2.0$~MK, respectively.  From the same data set, we have obtained $n_\mathrm{e} = 4.02 \times 10^9$~cm$^{-3}$ and $T_\mathrm{e} = 2.14$~MK.  The background magnetic field, $B_\mathrm{e}$, with which the pressure balance equation~(\ref{eq:pbeq}) is satisfied, is equal to $6$~G.  For the rotating macrospicule we assume that $n_\mathrm{i} = 1.0 \times 10^{10}$~cm$^{-3}$ and $n_\mathrm{e} = 1.0 \times 10^{9}$~cm$^{-3}$ to have at least one order denser jet with respect to the surrounding plasma.  Our choice for macrospicule temperature is $T_\mathrm{i} = 5.0 \times 10^5$~K, while that of its environment is supposed to be $T_\mathrm{e} = 1.0 \times 10^6$~K.  The external magnetic field, $B_\mathrm{e}$, was taken as $5$~G.  All aforementioned physical parameters of the jets are summarized in \textbf{Table~1}.  The plasma beta was calculated using $(6/5)c_\mathrm{s}^2/v_\mathrm{A}^2$, where $c_\mathrm{s} = (\gamma k_\mathrm{B}T/m_\mathrm{ion})^{1/2}$ is the sound speed (in which $\gamma = 5/3$, $k_\mathrm{B}$ is the Boltzmann's constant, $T$ the electron temperature, and $m_\mathrm{ion}$ the ion or proton mass), and $v_\mathrm{A} = B/(\mu n_\mathrm{ion}m_\mathrm{ion})^{1/2}$ is the Alfv\'en speed, in which expression $B$ is the full magnetic field ${=}(B_\phi^2 + B_z^2)^{1/2}$, and $n_\mathrm{ion}$ is the ion or proton number density.

\section{Wave Dispersion Relation}
\label{sec:dispeq}
A dispersion relation for the propagation of high-mode ($m \geqslant 2$) MHD waves in a magnetized axially moving and rotating twisted jet was derived by \citet{Zaqarashvili2015} and \citet{Cheremnykh2018b}.  That equation was obtained, however, under the assumption that both media (the jet and its environment) are incompressible plasmas.  As seen from the last column in \textbf{Table~1}, plasma beta is greater than $1$ in the first, third, and forth jets which implies that the plasma of each of the aforementioned jets can be considered as a nearly incompressible fluid \citep{Zank1993}.  It is seen from the same table that the plasma beta of the second jet is less than one as is in each of the jet environments and that is why it is reasonable to treat them as cool media.  Thus, the wave dispersion relation, derived, for instance, in \citealp{Zaqarashvili2015}, has to be modified.  In fact, we need two modified versions: one for the incompressible jet--cool environment configuration, and other for the cool jet--cool environment configuration.  We are not going to present in details the derivation of the modified dispersion equations on the basis of the governing MHD equations, but will only sketch the essential steps in that procedure.  The main philosophy in deriving the wave dispersion equation is to find solutions for the total pressure perturbation, $p_\mathrm{tot}$, and for the radial component, $\xi_r$ of the Lagrangian displacement, $\bm{\xi}$, and merge them at the tube perturbed boundary through the boundary conditions for their ($p_\mathrm{tot}$ and $\xi_r$) continuity \citep{Chandrasekhar1961}.  In the case of the first configuration, we start with the linearized ideal MHD equations, governing the incompressible dynamics of the perturbations in the spinning jet
\begin{equation}
\label{eq:momentum}
    \frac{\partial}{\partial t}\boldsymbol{v} + (\boldsymbol{U}\cdot \nabla)\boldsymbol{v} + (\boldsymbol{v}\cdot \nabla) \boldsymbol{U} = -\frac{\nabla p_\mathrm{tot}}{\rho_\mathrm{i}} + \frac{\left( \boldsymbol{B}_\mathrm{i} \cdot \nabla \right)\boldsymbol{b}}{\rho_\mathrm{i} \mu} + \frac{\left( \boldsymbol{b} \cdot \nabla \right)\boldsymbol{B}_\mathrm{i}}{\rho_\mathrm{i} \mu},
\end{equation}
\begin{equation}
\label{eq:induct}
	\frac{\partial}{\partial t}\boldsymbol{b} - \nabla \times \left( \boldsymbol{v}
    \times \boldsymbol{B}_\mathrm{i} \right) - \nabla \times \left( \boldsymbol{U} \times \boldsymbol{b}
    \right) = 0,	
\end{equation}
\begin{equation}
\label{eq:divv}
	\nabla \cdot \boldsymbol{v} = 0,	
\end{equation}
\begin{equation}
\label{eq:divb}
	\nabla \cdot \boldsymbol{b} = 0,
\end{equation}
where $\boldsymbol{v} = (v_r, v_\phi, v_z)$ and $\boldsymbol{b} = (b_r, b_\phi, b_z)$ are the perturbations of fluid velocity and magnetic field, respectively, and $p_\mathrm{tot}$ is the perturbation of the total pressure, $p_\mathrm{t} = p_\mathrm{i} + B_\mathrm{i}^2/2\mu$.  The Lagrangian displacement, $\bm{\xi}$, can be found from the fluid velocity perturbation, $\boldsymbol{v}$, using the relation \citep{Chandrasekhar1961}
\begin{equation}
\label{eq:v-xi}
    \boldsymbol{v} = \frac{\partial \bm{\xi}}{\partial t} + \left( \boldsymbol{U}\cdot \nabla \right)\bm{\xi} - \left( \boldsymbol{\bm{\xi}}\cdot \nabla \right)\boldsymbol{U}.
\end{equation}
Further on, assuming that all perturbations are ${\propto}\exp \left[\mathrm{i} \left( -\omega t + m \phi + k_z z \right) \right]$ and considering that the rotation and the magnetic field twists in the jet are uniform, that is,
\begin{equation}
\label{eq:uniform}
    U_\phi(r) = \Omega r \qquad  \mbox{and} \qquad B_{\mathrm{i}\phi}(r) = A r,
\end{equation}
where $\Omega$ and $A$ are constants, from the above set of equations, (\ref{eq:momentum})--(\ref{eq:v-xi}), we obtain the following dispersion equation of the MHD wave with mode number $m$ (for details see \citealp{Zaqarashvili2015}):
\begin{eqnarray}
\label{eq:dispeq}
    \frac{\left( \sigma^2 - \omega_\mathrm{Ai}^2 \right)F_m(\kappa_\mathrm{i}a) - 2m\left( \sigma \Omega + A\omega_\mathrm{Ai}/\! \sqrt{\mu \rho_\mathrm{i}} \right)}{\rho_\mathrm{i}\left( \sigma^2 - \omega_\mathrm{Ai}^2 \right)^2 - 4\rho_\mathrm{i}\left( \sigma \Omega + A\omega_\mathrm{Ai}/\! \sqrt{\mu \rho_\mathrm{i}} \right)^2} \nonumber \\
    \nonumber \\
    {}= \frac{P_m(\kappa_\mathrm{e} a)}{\rho_\mathrm{e}\left( \omega^2 - \omega_\mathrm{Ae}^2 \right) - \left( \rho_\mathrm{i}\Omega^2 - A^2/\mu \right)P_m(\kappa_\mathrm{e} a)},
\end{eqnarray}
where
\[
    F_m(\kappa_\mathrm{i}a) = \frac{\kappa_\mathrm{i}aI_m^{\prime}(\kappa_\mathrm{i}a)}{I_m(\kappa_\mathrm{i}a)} \quad \mbox{and} \quad P_m(\kappa_\mathrm{e} a) = \frac{\kappa_\mathrm{e} aK_m^{\prime}(\kappa_\mathrm{e} a)}{K_m(\kappa_\mathrm{e} a)}.
\]
In above expressions, the prime means differentiation of the Bessel functions with respect to their arguments,
\[
    \kappa_\mathrm{i}^2 = k_z^2\left[ 1 - 4\left( \frac{\sigma \Omega + A\omega_\mathrm{Ai}/\!\sqrt{\mu \rho_\mathrm{i}}}{\sigma^2 - \omega_\mathrm{Ai}^2} \right)^2 \right] \quad \mbox{and} \quad \kappa_\mathrm{e}^2 = k_z^2 \left[ 1 - \left( \omega/\omega_\mathrm{Ae} \right)^2 \right]
\]
are the squared wave amplitude attenuation coefficients in the jet and its environment, in which
\[
    \omega_\mathrm{Ai} = \left( \frac{m}{r}B_{\mathrm{i}\phi} + k_z B_{\mathrm{i}z} \right)/\sqrt{\mu \rho_\mathrm{i}} \quad \mbox{and} \quad \omega_\mathrm{Ae} = k_z B_\mathrm{e}/\sqrt{\mu \rho_\mathrm{e}}
\]
are the local Alfv\'en frequencies in both media, and
\[
    \sigma = \omega - \frac{m}{r}U_{\phi} - k_z U_z
\]
is the Doppler-shifted angular wave frequency in the jet.  We note that in the case of incompressible coronal plasma \citep{Zaqarashvili2015}, $\kappa_\mathrm{e} = k_z$, because at an incompressible environment the argument of the modified Bessel function of second kind, $K_m$, and its derivative, $K_m^\prime$, is $k_z a$.

The basic MHD equations for an ideal cool plasma are, generally, the same as the set of equations (\ref{eq:momentum})--(\ref{eq:v-xi}) with equation~(\ref{eq:divv}) replaced by the continuity equation
\begin{eqnarray*}
\label{eq:cont}
    \frac{\partial \rho_1}{\partial t} = -\nabla \cdot (\rho_0 \boldsymbol{v}_1 + \rho_1 \boldsymbol{U}) = 0.
\end{eqnarray*}
Recall that for cold plasmas the total pressure reduces to the magnetic pressure only, that is $p_\mathrm{t} = B_1^2/2\mu$, the $z$ component of the velocity perturbation is zero, \emph{i.e.}, $\boldsymbol{v}_1 = (v_{1r}, v_{1\phi}, 0)$, while $\boldsymbol{B}_1 = (B_{1r}, B_{1\phi}, B_{1z})$.  The above equation, which defines the density perturbation, is not used in the derivation of the wave dispersion relation because we are studying the propagation and stability of Alfv\'en-wave-like perturbations of the fluid velocity and magnetic field.  Following the standard scenario for deriving the MHD wave dispersion relation \citep{Zhelyazkov2018b}, we finally arrive at:
\begin{eqnarray}
\label{eq:dispeqc}
    \frac{\left( \sigma^2 - \omega_\mathrm{Ai}^2 \right)F_m(\kappa^\mathrm{c}_\mathrm{i}a) - 2m\left( \sigma \Omega + A\omega_\mathrm{Ai}/\! \sqrt{\mu \rho_\mathrm{i}} \right)}{\rho_\mathrm{i}\left( \sigma^2 - \omega_\mathrm{Ai}^2 \right)^2 - 4\rho_\mathrm{i}\left( \sigma \Omega + A\omega_\mathrm{Ai}/\! \sqrt{\mu \rho_\mathrm{i}} \right)^2} \nonumber \\
    \nonumber \\
    {}= \frac{P_m(\kappa^\mathrm{c}_\mathrm{e} a)}{\rho_\mathrm{e}\left( \omega^2 - \omega_\mathrm{Ae}^2 \right) - \left( \rho_\mathrm{i}\Omega^2 - A^2/\mu \right)P_m(\kappa^\mathrm{c}_\mathrm{e} a)},
\end{eqnarray}
where
\[
    F_m(\kappa^\mathrm{c}_\mathrm{i}a) = \frac{\kappa^\mathrm{c}_\mathrm{i}aI_m^{\prime}(\kappa^\mathrm{c}_\mathrm{i}a)}{I_m(\kappa^\mathrm{c}_\mathrm{i}a)} \quad \mbox{and} \quad P_m(\kappa^\mathrm{c}_\mathrm{e} a) = \frac{\kappa^\mathrm{c}_\mathrm{e} aK_m^{\prime}(\kappa^\mathrm{c}_\mathrm{e} a)}{K_m(\kappa^\mathrm{c}_\mathrm{e} a)}.
\]
Here, the wave attenuation coefficient in the internal medium has the form
\[
    \kappa^\mathrm{c}_\mathrm{i} = k_z\left\{ 1 - 4\left( \frac{\sigma \Omega + A\omega_\mathrm{Ai}/\!\sqrt{\mu \rho_\mathrm{i}}}{\sigma^2 - \omega_\mathrm{Ai}^2} \right)^2 \right\}^{1/2}\left( 1 - \frac{\sigma^2}{\omega_\mathrm{Ai}^2}  \right)^{1/2},
\]
while that in the environment, with $\Omega = 0$ and $A = 0$, is given by
\[
    \kappa^\mathrm{c}_\mathrm{e} = k_z\left( 1 - \frac{\omega^2}{\omega_\mathrm{Ae}^2} \right)^{1/2}.
\]
Note that (i) both dispersion relations, (\ref{eq:dispeq}) and (\ref{eq:dispeqc}), have similar forms---the difference is in the expressions for the wave attenuation coefficient inside the jet, namely $\kappa^\mathrm{c}_\mathrm{i} = \kappa_\mathrm{i}\left[ 1 - \sigma^2/\omega_\mathrm{Ai}^2 \right]^{1/2}$; and (ii) the wave attenuation coefficients in the environments are not surprisingly the same, that is, $\kappa^\mathrm{c}_\mathrm{e} = \kappa_\mathrm{e} \equiv \left[ 1 - \omega^2/\omega_\mathrm{Ae}^2 \right]^{1/2}$.

\section{Numerical Solutions, Wave Dispersion, and Growth Rate Diagrams}
\label{sec:numerics}
In studying at which conditions the high ($m \geqslant 2$) MHD modes in a jet--coronal plasma system become unstable, that is, all the perturbations to grow exponentially in time, we have to consider the wave angular frequency, $\omega$, as a complex quantity: $\omega \equiv \mathrm{Re}(\omega) + \mathrm{i}\,\mathrm{Im}(\omega)$ in contrast to the wave mode number, $m$, and propagating wavenumber, $k_z$, which are real quantities.  The $\mathrm{Re}(\omega)$ is responsible for the wave dispersion while the $\mathrm{Im}(\omega)$ yields the wave growth rate.  In the numerical task for finding the complex solutions to the wave dispersion relation~(\ref{eq:dispeq}) or (\ref{eq:dispeqc}), it is convenient to normalize all velocities with respect to the Alfv\'en speed inside the jet, defined as $v_\mathrm{Ai} = B_{\mathrm{i}z}/\sqrt{\mu \rho_\mathrm{i}}$, and the lengths with respect to $a$.  Thus, we have to search the real and imaginary parts of the nondimensional wave phase velocity, $v_\mathrm{ph} = \omega/k_z$, that is, Re($v_\mathrm{ph}/v_\mathrm{Ai}$) and Im($v_\mathrm{ph}/v_\mathrm{Ai}$) as functions of the normalized wavenumber $k_z a$.  The normalization of the other quantities like the local Alfv\'en and Doppler-shifted frequencies alongside the Alfv\'en speed in the environment, $v_\mathrm{Ae} = B_\mathrm{e}/\sqrt{\mu \rho_\mathrm{e}}$, requires the usage of both twist parameters, $\varepsilon_1$ and $\varepsilon_2$, and also of the magnetic fields ratio, $b = B_\mathrm{e}/B_{\mathrm{i}z}$.  The nondimensional form of the jet axial velocity, $U_z$, is given by the Alfv\'en Mach number $M_\mathrm{A} = U_z/v_\mathrm{Ai}$.  Another important nondimensional  parameter is the density contrast between the jet and its surrounding medium, $\eta = \rho_\mathrm{e}/\rho_\mathrm{i}$.  Hence, the input parameters in the numerical task of finding the solutions to the transcendental equation~(\ref{eq:dispeq}) or (\ref{eq:dispeqc}) (in complex variables) are: $m$, $\eta$, $\varepsilon_1$, $\varepsilon_2$, $b$, and $M_\mathrm{A}$.  \citet{Zaqarashvili2015} have established that KHI in an untwisted $(A = 0)$ rotating flux tube with negligible longitudinal velocity can occur if
\begin{equation}
\label{eq:criterion}
    \frac{a^2 \Omega^2}{v_\mathrm{Ai}^2} > \frac{1 + \eta}{1 + |m|\eta}\,\frac{(k_z a)^2}{|m| - 1}(1 + b^2).
\end{equation}
This inequality says that each MHD wave with mode number $m \geqslant 2$, propagating in a rotating jet can become unstable.  This instability condition can be used also in the cases of slightly twisted spinning jets, provided that the magnetic field twist parameter, $\varepsilon_1$, is a number lying in the range of $0.001$--$0.005$, simply because the numerical solutions, for example, to equation (\ref{eq:dispeq}) show that practically there is no difference between the instability ranges at $\varepsilon_1 = 0$, and at $0.001$ or $0.005$.  An important step in our study is
the supposition that the deduced from observations jet axial velocity, $U_z$, is the threshold speed for the KHI occurrence.  Then, for fixed values of $m$, $\eta$, $U_\phi = \Omega a$, $v_\mathrm{Ai}$, and $b$, the inequality~(\ref{eq:criterion}) can be rearranged to define the upper limit of the instability range on the $k_z a$-axis
\begin{equation}
\label{eq:instcond}
    (k_z a)_\mathrm{rhs} < \left\{ \left( \frac{U_\phi}{v_\mathrm{Ai}} \right)^2 \frac{1 + |m|\eta}{1 + \eta}\,\frac{|m| - 1}{1 + b^2} \right\}^{1/2}.
\end{equation}
According to the above inequality, the KHI can occur for nondimensional wavenumbers $k_z a$ less than $(k_z a)_\mathrm{rhs}$.  On the other hand, one can talk for instability if the unstable wavelength, $\lambda_\mathrm{KH} = 2\pi/k_z$, is shorter than the height of the jet, $H$, which means that the lower limit of the instability region is given by:
\begin{equation}
\label{eq:lhlimit}
    (k_z a)_\mathrm{lhs} > \frac{\pi \Delta \ell}{H},
\end{equation}
where $\Delta \ell$ is the jet width.  Hence, the instability range in the $k_z a$-space is $(k_z a)_\mathrm{lhs} < k_z a < (k_z a)_\mathrm{rhs}$.  Note that the lower limit, $(k_z a)_\mathrm{lhs}$, is fixed by the width and height of the jet, while the upper limit, $(k_z a)_\mathrm{rhs}$, depends on several jet--environment parameters.  At fixed $U_\phi$,  $v_\mathrm{Ai}$, $\eta$, and $b$, the $(k_z a)_\mathrm{rhs}$ is determined by the MHD wave mode number, $|m|$.  As seen from inequality~(\ref{eq:instcond}), with increasing the $m$, that limit shifts to the right, that is, the instability range becomes wider.  The numerical solutions to the wave dispersion relation (\ref{eq:dispeq}) confirm this and for given $m$ one can obtain a series of unstable wavelengths, $\lambda_\mathrm{KH} = \pi\,\Delta \ell/k_z a$, as the shortest one takes place at $k_z a \approx (k_z a)_\mathrm{rhs}$.  For relatively small mode numbers, when even the shortest unstable wavelengths turn out to be a few tens megameters, that could hardly be associated with the observed KH ones.  As observations show, the KHI vortex-like structures running at the boundary of the jet, have the size of the width or radius of the flux tube (see, for instance, Fig.~1 in \citealp{Zhelyazkov2018a}).  Therefore, we have to look for such an $m$, whose instability range would accommodate the expected unstable wavelength presented by its nondimensional wavenumber, $k_z a = \pi \Delta \ell/\lambda_\mathrm{KH}$.  An estimation of the required mode number for an $\varepsilon_1 = 0.005$-rotating flux tube can be obtained by presenting the instability criterion (\ref{eq:criterion}) in the form
\begin{equation}
\label{eq:findingm}
    \eta |m|^2 + (1 - \eta)|m| - 1 - \frac{(k_z a)^2(1 + \eta)(1 + b^2)}{(U_\phi/v_\mathrm{Ai})^2} > 0.
\end{equation}
We will use this inequality for obtaining the optimal $m$ for each of the studied jets by specifying the value of that $k_z a$ (along with the other aforementioned input parameters) which corresponds to the expected unstable wavelength $\lambda_\mathrm{KH}$.

\subsection{Kelvin--Helmholtz Instability in a Standard Polar Coronal Hole Jet}
\label{subsec:coronal_hole}
\begin{figure}[!ht]
   \centering
   \includegraphics[height=.30\textheight]{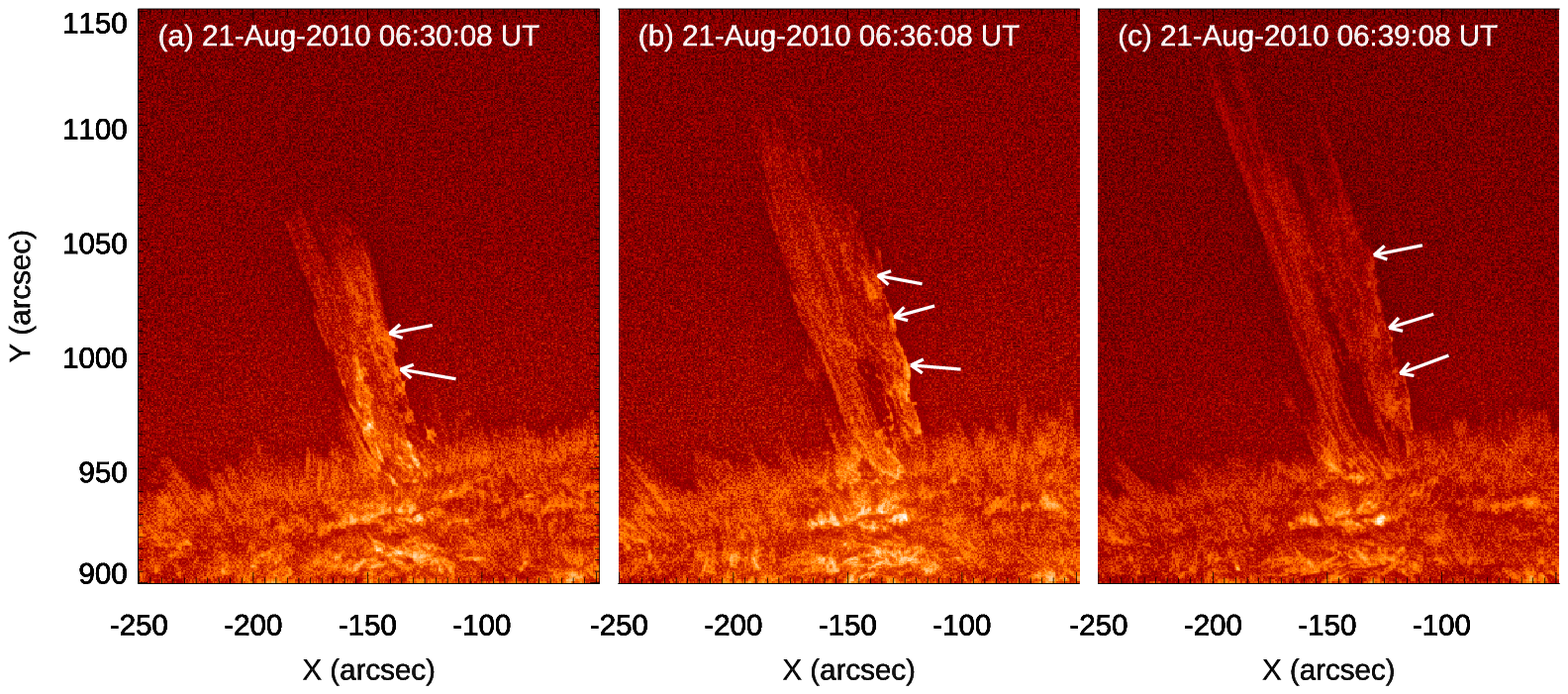}
   \caption{AIA $304$~\AA\ images showing the detailed evolution of the jet observed on 2010 August 21.  The small moving blobs on the right side boundary of the jet as indicated by white arrows, could be produced by a KHI.}
   \label{fig:fig1}
\end{figure}

\citet{Chen2012} observationally studied the jet event of 2010 August 21, which occurred in the coronal hole region, close to the north pole of the Sun.  \textbf{Figure~\ref{fig:fig1}} presents the jet's evolution in AIA $304$~\AA.  The jet started around  06:07~\textsc{ut}, reached its maximum height around 06:40~\textsc{ut}.  During the evolution of the jet between 06:32 and 06:38~\textsc{ut}, small scale moving blobs appeared on the right boundary.  We interpret these blobs, shown by arrows in \textbf{Figure~\ref{fig:fig1}}, as evidence of KHI.  By tracking six identified moving features in the jet, \citet{Chen2012} found that the plasma moved at an approximately constant speed along the jet's axis.  Inferred from linear and trigonometric fittings to the axial and transverse heights of the six tracks, the authors have found that the mean values of the axial velocity, $U_z$, transfer/rotational velocity, $U_\phi$, angular speed, $\Omega$, rotation period, $T$, and rotation radius, $a$, are $114$~km\,s$^{-1}$, $136$~km\,s$^{-1}$, $0.81^\circ$s$^{-1}$ (or $14.1 \times 10^{-3}$~rad\,s$^{-1}$), $452$~s, and $9.8 \times 10^3$~km, respectively.  The height of the jet is evaluated as $H = 179$~Mm.

\begin{figure}[!ht]
   \centerline{\hspace*{0.015\textwidth}
               \includegraphics[width=0.515\textwidth,clip=]{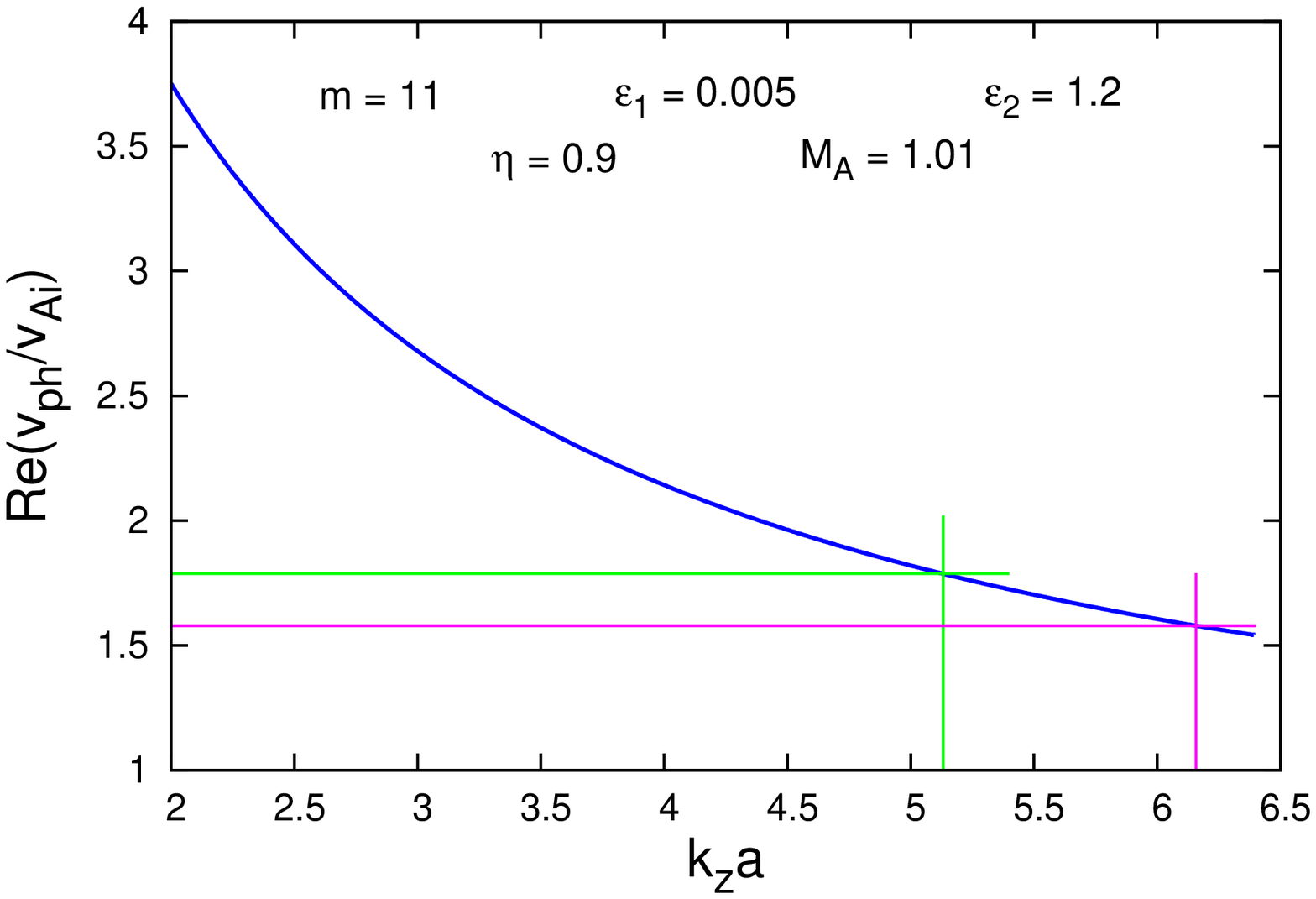}
               \hspace*{-0.03\textwidth}
               \includegraphics[width=0.515\textwidth,clip=]{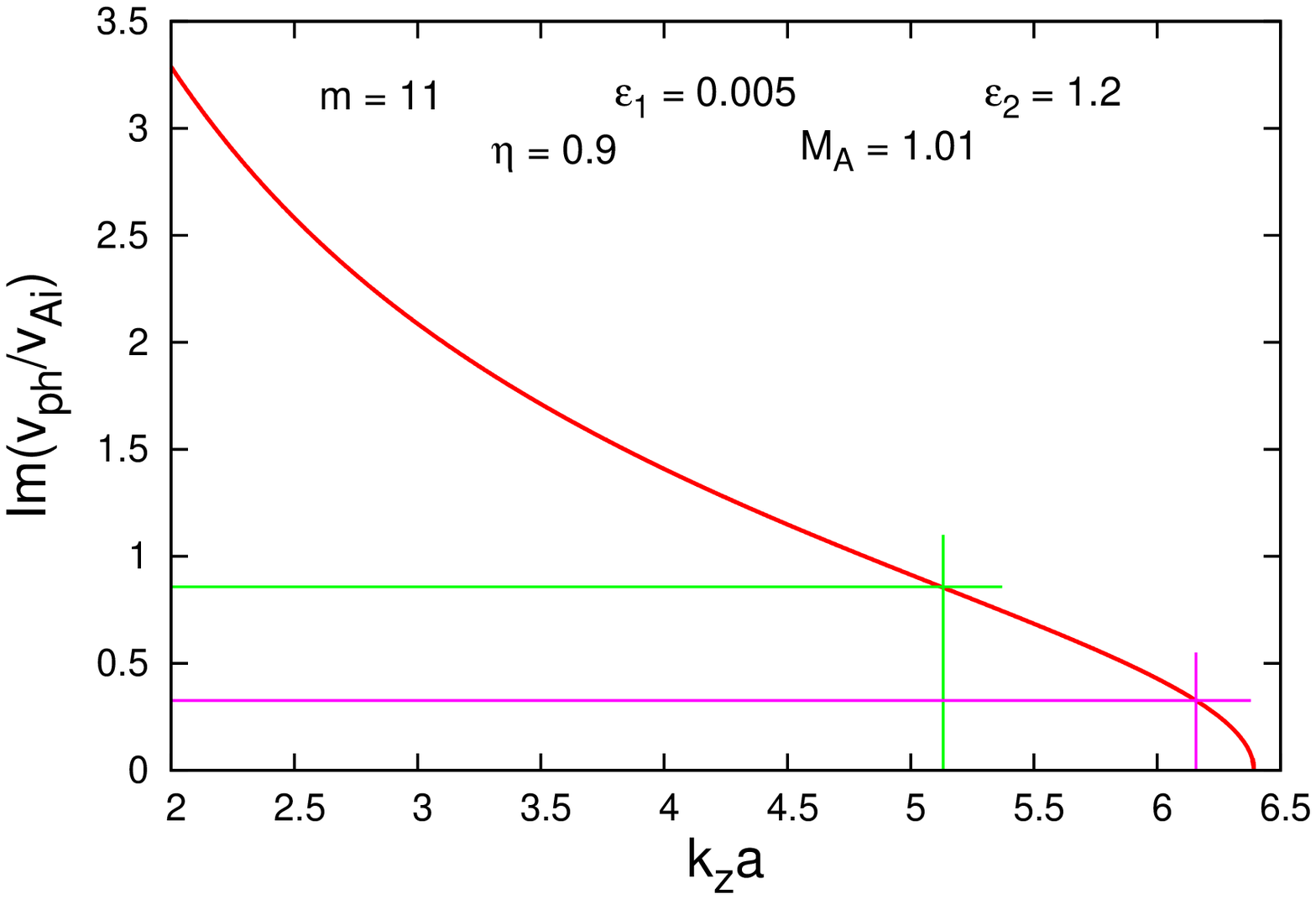}
              }
  \caption{(Left) Dispersion curve of the $m = 11$ MHD mode propagating along a twisted incompressible coronal hole jet at $\eta = 0.9$, $b = 1.834$, $M_{\rm A} = 1.01$, $\varepsilon_1 = 0.005$, and $\varepsilon_2 = 1.2$.  (Right) Normalized growth rate curve of the $m = 11$ MHD mode computed at the same input parameters as in the left panel.  The crosses of purple and green lines yield the normalized values of the wave phase velocity and the wave growth rate at the two unstable wavelengths of $10$ and $12$~Mm, respectively.}
   \label{fig:fig2}
\end{figure}

It seems reasonable the shortest unstable wavelength, $\lambda_\mathrm{KH}$, to be equal to $10$~Mm (approximately half of the jet width, $\Delta \ell = 19.6$~Mm), which implies that its position on the $k_z a$ one-dimensional space is $k_z a = 6.158$.  The input parameters, necessary to find out that MHD wave mode number, whose instability range will contain the nondimensional wavenumber of $6.158$, using inequality (\ref{eq:findingm}), are accordingly (see \textbf{Table~1}) $\eta = 0.9$, $b = 1.834$, $v_\mathrm{Ai} = 112.75$~km\,s$^{-1}$, and $U_\phi = 136$~km\,s$^{-1}$.  (We note, that the values of $b$ and $v_\mathrm{Ai}$ were obtained with the help of equation~(\ref{eq:pbeq}) assuming that the magnetic field twist is $\varepsilon_1 = 0.005$.)  With these entry data, from inequality (\ref{eq:findingm}) one obtains that $|m| > 15$ should provide the required instability region or window.  The numerical solutions to equation~(\ref{eq:dispeq}) show that this value of $m$ is overestimated---an $m = 11$ turns out to be perfect for the case.  The discrepancy between the predicted and computed value of $|m|$ is not surprising because inequality (\ref{eq:findingm}) yields only an indicative value.  The input parameters for finding the solutions to the dispersion equation~(\ref{eq:dispeq}) are as follows: $m = 11$, $\eta = 0.9$, $\varepsilon_1 = 0.005$, $\varepsilon_2 = 1.2$, $b = 1.834$, and $M_\mathrm{A} = 1.01$ $({=}114/112.75)$.  The results of computations are graphically presented in \textbf{Figure~\ref{fig:fig2}}.  From that figure, one can obtain the normalized wave phase velocity, $\mathrm{Re}(v_\mathrm{ph}/v_\mathrm{Ai})$, and the normalized growth rate, $\mathrm{Im}(v_\mathrm{ph}/v_\mathrm{Ai})$, of the unstable $\lambda_\mathrm{KH} = 10$~Mm wave, both read at the purple cross points.  From the same plot, one can find the instability characteristics at another wavelength, precisely $\lambda_\mathrm{KH} = 12$~Mm, whose position on the $k_z a$-axis is fixed at $k_z a = 5.131$.  The values of nondimensional wave phase velocity and growth rate can be read from the green cross points.  The KHI wave growth rate, $\gamma_\mathrm{KH}$, growth time, $\tau_\mathrm{KH} = 2\pi/\gamma_\mathrm{KH}$, and wave velocity, $v_\mathrm{ph}$, in absolute units, estimated from the plots in \textbf{Figure~\ref{fig:fig2}}, for the two wavelengths, are
\begin{eqnarray*}
    \gamma_\mathrm{KH} \cong 23.09 \times 10^{-3}\:\mathrm{s}^{-1}, \;\, \tau_\mathrm{KH} \cong 4.5\:\mathrm{min}, \;\, v_\mathrm{ph} \cong 178\:\mathrm{km}\,\mathrm{s}^{-1},\;\,\mbox{for} \;\,\lambda_\mathrm{KH} = 10\:\mathrm{Mm},
\end{eqnarray*}
and
\begin{eqnarray*}
    \gamma_\mathrm{KH} \cong 50.65 \times 10^{-3}\:\mathrm{s}^{-1}, \;\, \tau_\mathrm{KH} \cong 2.1\:\mathrm{min}, \;\, v_\mathrm{ph} \cong 202\:\mathrm{km}\,\mathrm{s}^{-1},\;\,\mbox{for}\;\,\lambda_\mathrm{KH} = 12\:\mathrm{Mm}.
\end{eqnarray*}

Let us recall that the value of the Alfv\'en speed used in the normalization is $v_\mathrm{Ai} = 112.75$~km\,s$^{-1}$.  We see that the two wave phase velocities are slightly super-Alfv\'enic and when moving along the $k_z a$-axis to the left, the normalized wave velocity becomes higher.  If we fix a $k_z a$-position near the lower limit of the unstable region, $(k_z a)_\mathrm{lhs} = 0.344$, say, at $k_z a = 0.513$, which means $\lambda_\mathrm{KH} = 120$~Mm, the KHI characteristics obtained from the numerical solutions to equation~(\ref{eq:dispeq}) are $\tau_\mathrm{KH} = 1.4$~min and $v_\mathrm{ph} = 1\,473$~km\,s$^{-1}$, respectively.  As we have discussed in \citealp{Zhelyazkov2018a}, ``the KHI growth time could be estimated from the temporal evolution of the blobs in their initial stage and it was found to be about $2$--$4$ minutes'', so the instability developing times of $2.1$ and $4.5$~min obtained from our plots are in good agreement with the observations.

\begin{figure}[!ht]
   \centerline{\hspace*{0.015\textwidth}
               \includegraphics[width=0.515\textwidth,clip=]{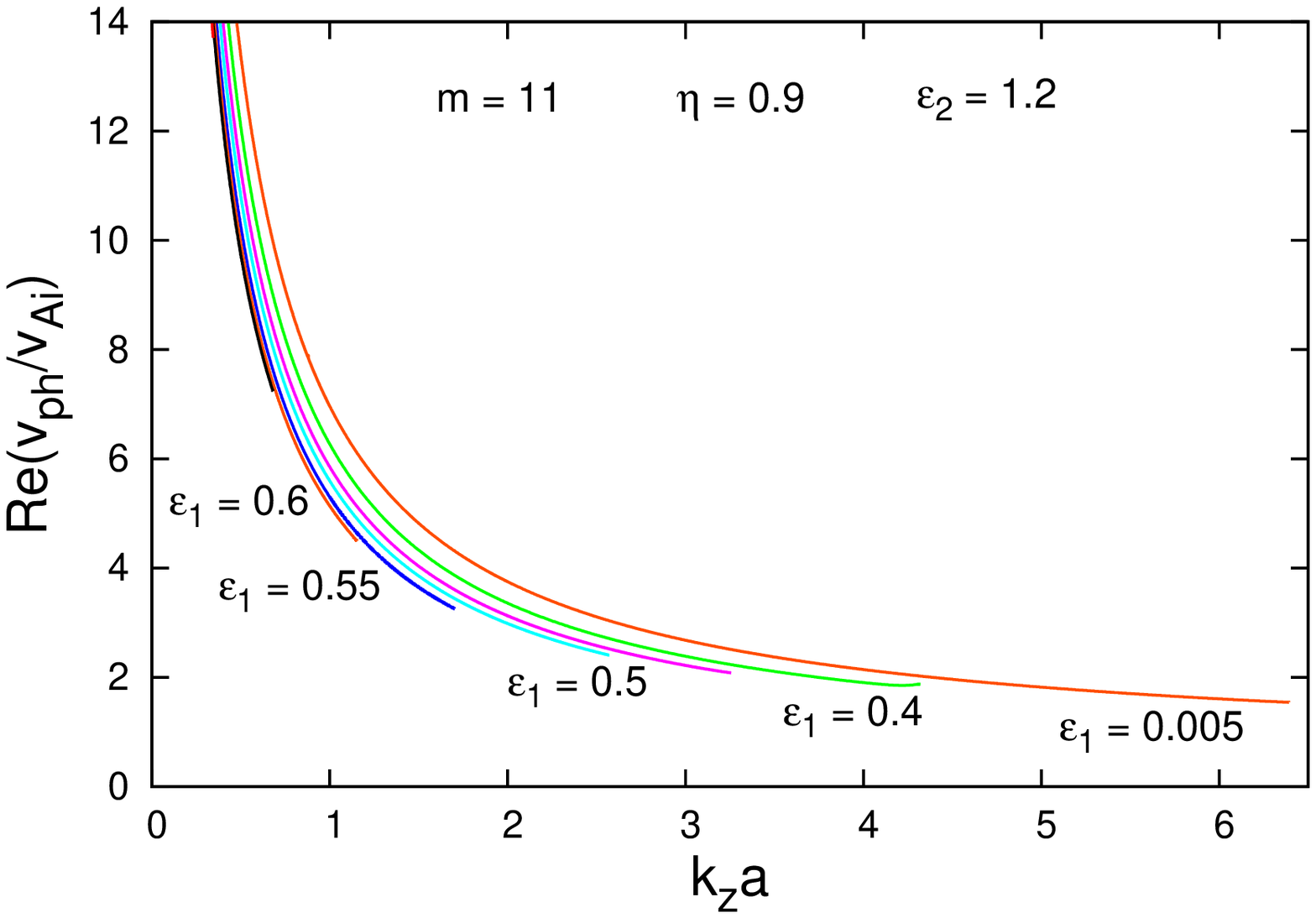}
               \hspace*{-0.03\textwidth}
               \includegraphics[width=0.515\textwidth,clip=]{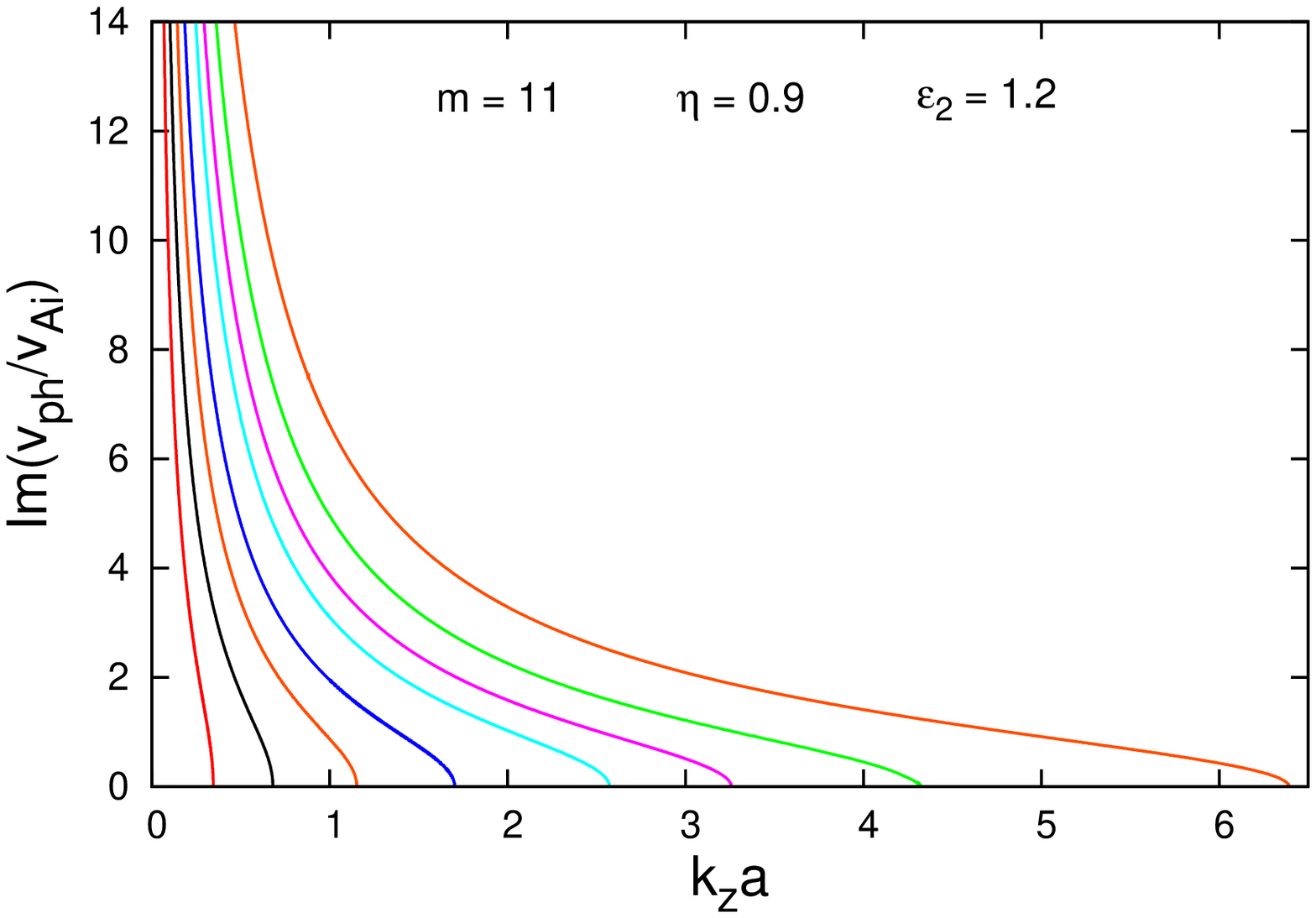}
              }
  \caption{(Left) Dispersion curves of the unstable $m = 11$ MHD mode propagating along a twisted incompressible jet in a coronal hole at $\eta = 0.9$, $\varepsilon_2 = 1.2$, and the following values of $\varepsilon_1$ (from right to left): $0.005$, $0.4$, $0.5$, $0.55$, $0.6$, $0.625$, $0.645$, and $0.653577$ (red curve in the right plot).  Alfv\'en Mach numbers for these curves are respectively $1.01$, $0.93$, $0.88$, $0.84$, $0.81$, $0.79$, $0.77$, and $0.7653$.  (Right) Growth rates of the unstable $m = 11$ mode for the same input parameters.  The azimuthal magnetic field that corresponds to $\varepsilon_1^\mathrm{cr} = 0.653577$ (the instability window with zero width) and stops the KHI onset is equal to $1.4$~G.}
   \label{fig:fig3}
\end{figure}

A specific property of the instability $k_z a$-ranges is that for a fixed mode number, $m$, their widths depend upon $\varepsilon_1$ and with increasing the value of $\varepsilon_1$, the instability window becomes narrower and at some critical $\varepsilon_1$ its width equals zero.  In our case that happens with $\varepsilon_1^\mathrm{cr} = 0.653577$ at $(k_z a)_\mathrm{lhs} = 0.344$.  In \textbf{Figure~\ref{fig:fig3}}, curves of dimensionless $v_\mathrm{ph}$ and $\gamma_\mathrm{KH}$ have been plotted for several $\varepsilon_1$ values.  Note that each larger value of $\varepsilon_1$ implies an increase in $B_{\mathrm{i}\phi}$.  But that increase in $B_{\mathrm{i}\phi}$ requires an increase in $B_{\mathrm{i}z}$ too, in order the total pressure balance equation~(\ref{eq:pbeq}) to be satisfied under the condition that the hydrodynamic pressure term and the environment total pressure are fixed.  The increase in $B_{\mathrm{i}z}$ (and in the full magnetic field $B_\mathrm{i}$) implies a decrease both in the magnetic field ratio, $b$, and in the Alfv\'en Mach number, $M_\mathrm{A}$.  Thus, gradually increasing the magnetic field twist $\varepsilon_1$ from $0.005$ to $0.653577$, we get a series of dispersion and growth rate curves with progressively diminishing parameters $b$ and $M_\mathrm{A}$.  The red growth rate curve in the right panel of \textbf{Figure~\ref{fig:fig3}} has been obtained for $\varepsilon_1^\mathrm{cr} = 0.653577$ with $M_\mathrm{A} = 0.7652$ and it visually fixes the lower limit of all other instability windows.  The azimuthal magnetic field $B_{\mathrm{i}\phi}^\mathrm{cr}$ that stops the KHI, computed at $B_\mathrm{i} = 2.58$~G, is equal to $1.4$~G.

\subsection{Kelvin--Helmholtz Instability in a Blowout Polar Coronal Hole Jet}
\label{subsec:blowout_hole}
\begin{figure}[!ht]
   \centering
   \includegraphics[height=.495\textheight,clip=]{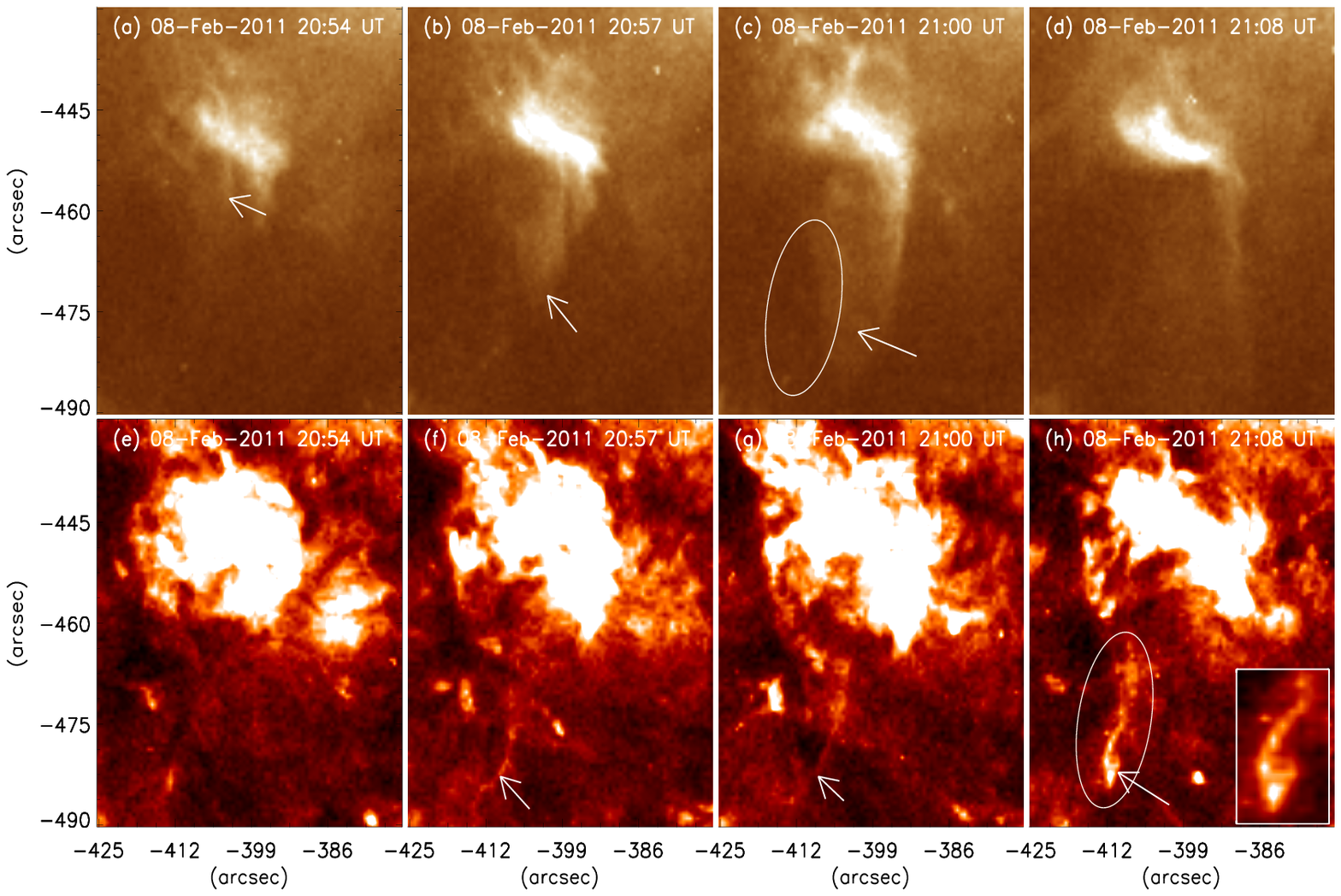}
   \caption{AIA $193$ top \textsf{\textbf{(a--d)}} and $304$~\AA{} bottom \textsf{\textbf{(e--h)}} images showing the evolution of the jet observed on 2011 February 8 marked by white arrows could be produced by KHI.  For the better visibility of the jet, we have saturated the image at the foot-point.  The ellipse in the \textsf{\textbf{(c--h)}} represents the east edge of the jets.  In $304$~\AA{} at ${\sim}$21:08~\textsc{ut} we have observed the blob structures at the eastern boundary.  The enlarged view of the blobs is shown in the inset in \textsf{\textbf{(h)}}.}
   \label{fig:fig4}
\end{figure}

\citet{Young2014a} observed a small blowout jet at the boundary of the south polar coronal hole on 2011 February 8 at around 21:00~\textsc{ut}.  The evolution of jet observed by the AIA is displayed in \textbf{Figure~\ref{fig:fig4}}.  The jet activity was between 20:50 to 21:15~\textsc{ut}.  This coronal hole is centered around  $x = -400$~arcsec, $y = -400$~arcsec.  The jet has very broad and faint structure and is ejected in the south direction.   We could see the evolution of jet in the AIA $193$~\AA\ clearly. However, in AIA $304$~\AA\, the whole jet is not visible.  Moreover, we observe the eastern boundary of the jet in AIA $304$~\AA{}.  During its evolution in $304$~\AA\, we found the blob structures at the jet boundary.  These blobs could be due to the KHI as reported in previous observations (see, for example \citealp{Zhelyazkov2018a}).  At the jet initiation/base site, we observed the coronal hole bright points. These bright points are the results of coronal low-laying loops reconnection \citep{Madjarska2019}.

\begin{figure}[!ht]
   \centerline{\hspace*{0.015\textwidth}
               \includegraphics[width=0.515\textwidth,clip=]{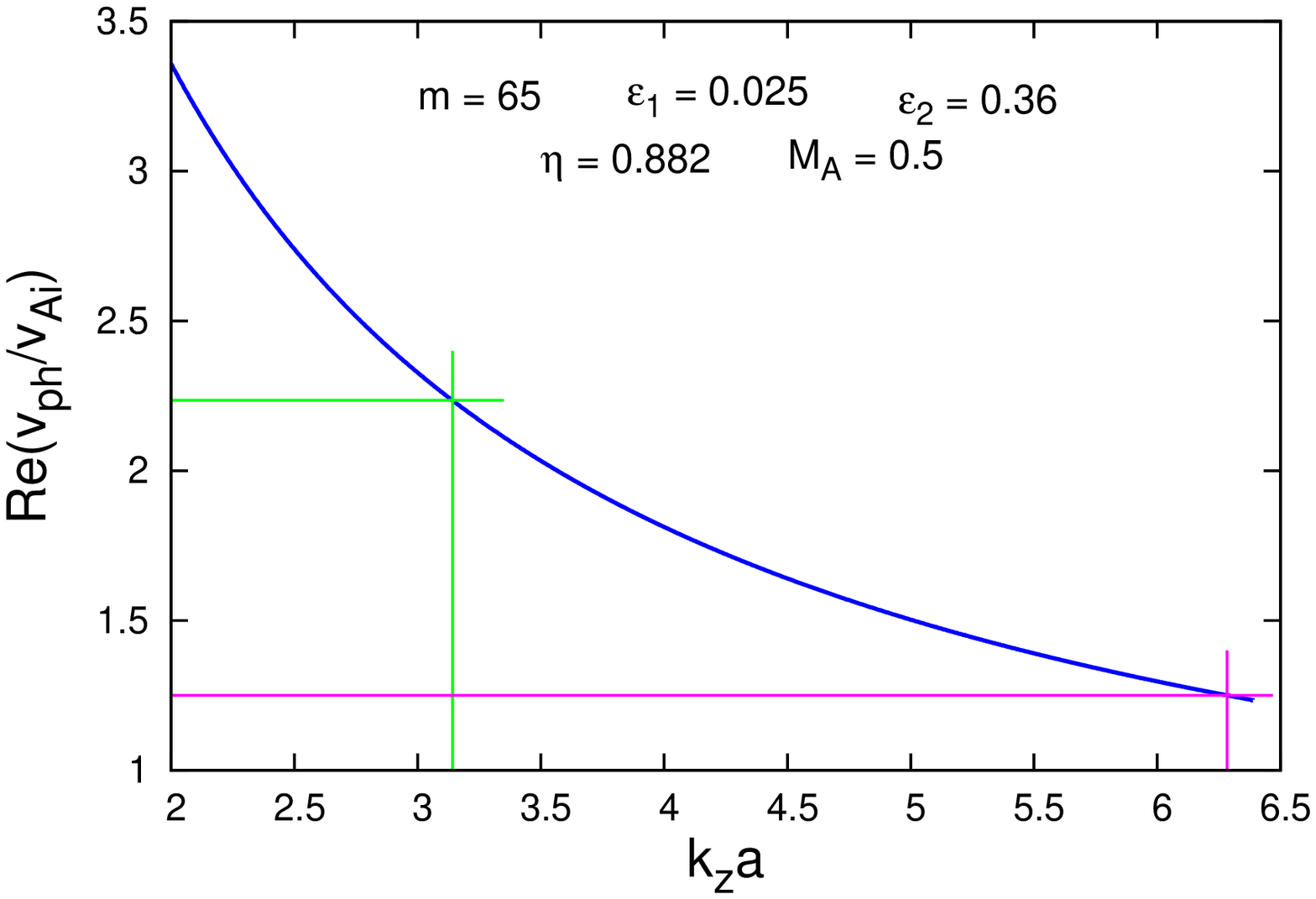}
               \hspace*{-0.03\textwidth}
               \includegraphics[width=0.515\textwidth,clip=]{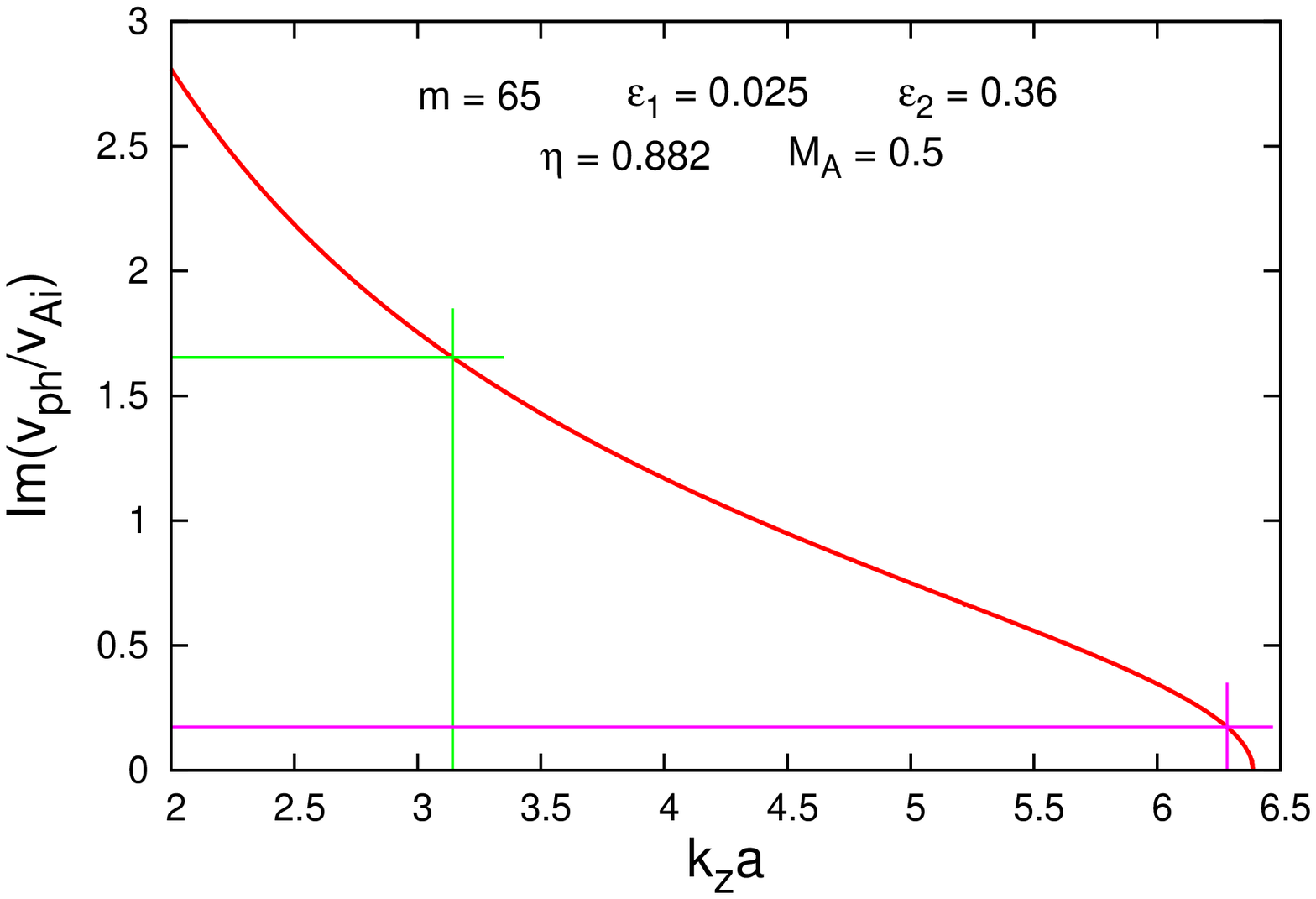}
              }
  \caption{(Left) Dispersion curve of the $m = 65$ MHD mode propagating along a twisted cool coronal hole jet at $\eta = 0.882$, $b = 1.014$, $M_{\rm A} = 0.5$, $\varepsilon_1 = 0.025$, and $\varepsilon_2 = 0.36$.  (Right) Normalized growth rate curve of the $m = 65$ MHD mode computed at the same input parameters as in the left panel.  The crosses of purple and green lines yield the normalized values of the wave phase velocity and the wave growth rate at the two unstable wavelengths of $7.5$ and $15$~Mm, respectively.}
   \label{fig:fig5}
\end{figure}

According to \citet{Young2014a} estimations, the jet is extended for $H = 30$~Mm with a width of $\Delta \ell = 15$~Mm.  The jet duration is $25$~min and the bright point is not significantly disrupted by the jet occurrence.  The jet $n$ is $n_\mathrm{i} = 1.7 \times 10^8$~cm$^{-3}$, while that of the surrounding coronal plasma we assume to be $n_\mathrm{e} = 1.5 \times 10^8$~cm$^{-3}$.  The jet temperature is $T_\mathrm{i}= 1.7$~MK and the environment one is $T_\mathrm{e}= 2.0$~MK.  The jet axial velocity is $U_z = 250$~km\,s$^{-1}$ and the rotational one is $U_\phi = 90$~km\,s$^{-1}$.  Assuming a magnetic field twist $\varepsilon_1 = 0.025$ and $B_\mathrm{e} = 3$~G, from equation~(\ref{eq:pbeq}), we obtain $\eta = 0.882$, $v_\mathrm{Ai} = 494.7$~km\,s$^{-1}$ (Alfv\'en speed in the environment is $v_\mathrm{Ae} = 534.0$~km\,s$^{-1}$), and $b = 1.014$.  We note, that while in the derivation of  equation~(\ref{eq:dispeqc}) we have neglected the thermal pressures, here, in using equation~(\ref{eq:pbeq}), we kept them.  If we anticipate that the shortest unstable wavelength is equal to $7.5$~Mm (with $k_z a = 2\pi$), the mode number $m$ whose instability range would accommodate the aforementioned wavelength, according to inequality~(\ref{eq:findingm}) must be at least $|m| = 71$.  The numerics show that the suitable $m$ is $|m| = 65$.  Thus, the input parameters for obtaining the numerical solutions to equation~(\ref{eq:dispeqc}) are: $m = 65$, $\eta= 0.882$, $\varepsilon_1 = 0.025$, $\varepsilon_2 = 0.36$ (${=}90/250$), $b = 1.014$, and $M_\mathrm{A} = 0.505$ (${\cong}250.0/494.7$).  The results are illustrated in \textbf{Figure~\ref{fig:fig5}}.  Along with $\lambda_\mathrm{KH} = 7.5$~Mm (purple lines), we have calculated the KHI characteristics also for $\lambda_\mathrm{KH} = 15$~Mm (at $k_z a = \pi$) (green lines), and they are
\begin{eqnarray*}
    \gamma_\mathrm{KH} = 72.0 \times 10^{-3}\:\mathrm{s}^{-1}, \;\, \tau_\mathrm{KH} \cong 1.5\:\mathrm{min}, \;\, v_\mathrm{ph} \cong 618\:\mathrm{km}\,\mathrm{s}^{-1},\;\,\mbox{for} \;\,\lambda_\mathrm{KH} = 7.5\:\mathrm{Mm},
\end{eqnarray*}
and
\begin{eqnarray*}
    \gamma_\mathrm{KH} \cong 342.75 \times 10^{-3}\:\mathrm{s}^{-1}, \;\, \tau_\mathrm{KH} \cong 0.3\:\mathrm{min}, \;\, v_\mathrm{ph} \cong 1106\:\mathrm{km}\,\mathrm{s}^{-1},\;\,\mbox{for}\;\,\lambda_\mathrm{KH} = 15\:\mathrm{Mm}.
\end{eqnarray*}
It is seen from the left panel that the unstable $m = 65$ MHD waves are generally super-Alfv\'enic. Since the instability developing times of the $m = 65$ mode are relatively short, that is, much less than the jet lifetime of $25$~min, we can conclude that the KHI in this jet is relatively fast.

\begin{figure}[!ht]
   \centerline{\hspace*{0.015\textwidth}
               \includegraphics[width=0.515\textwidth,clip=]{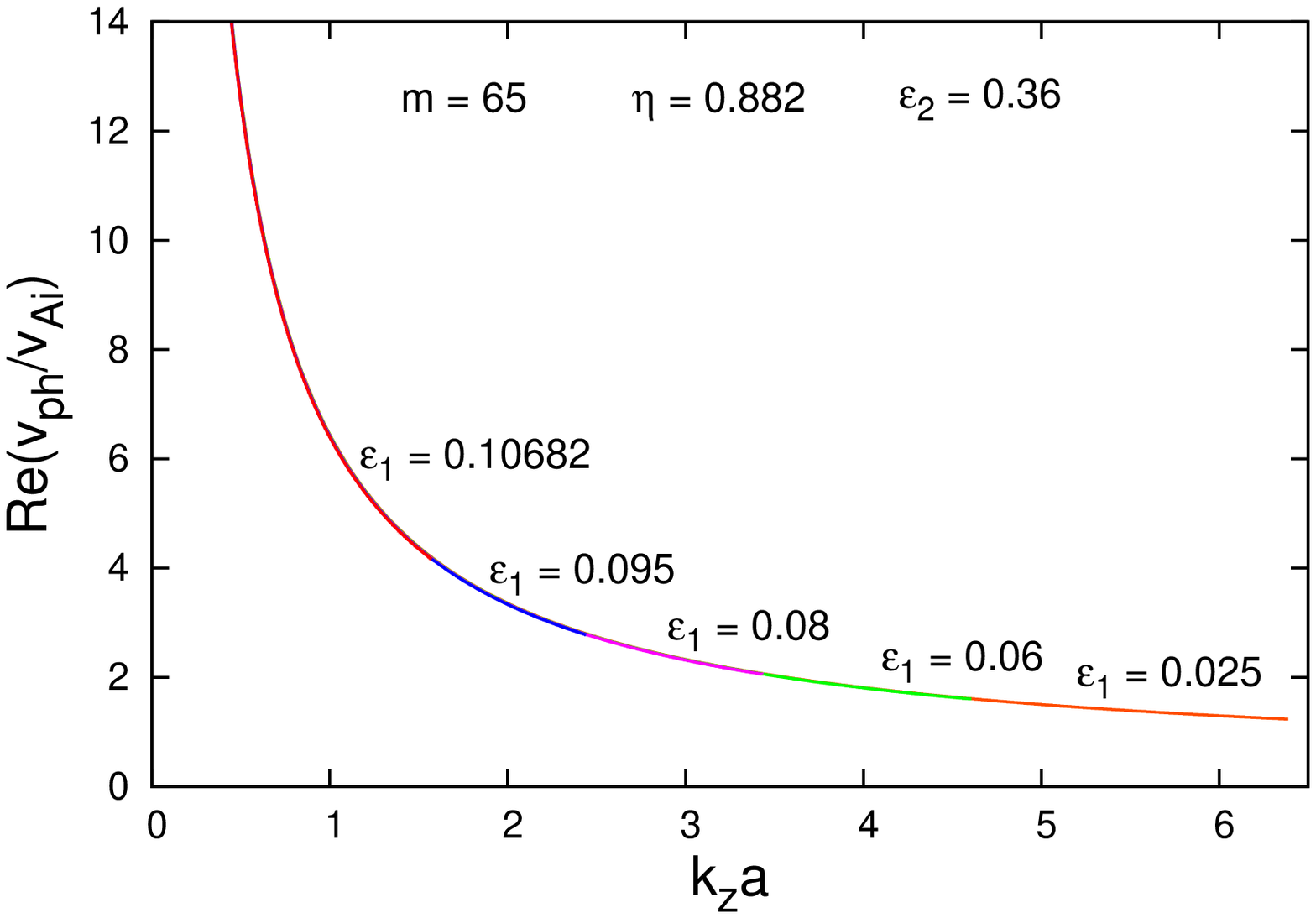}
               \hspace*{-0.03\textwidth}
               \includegraphics[width=0.515\textwidth,clip=]{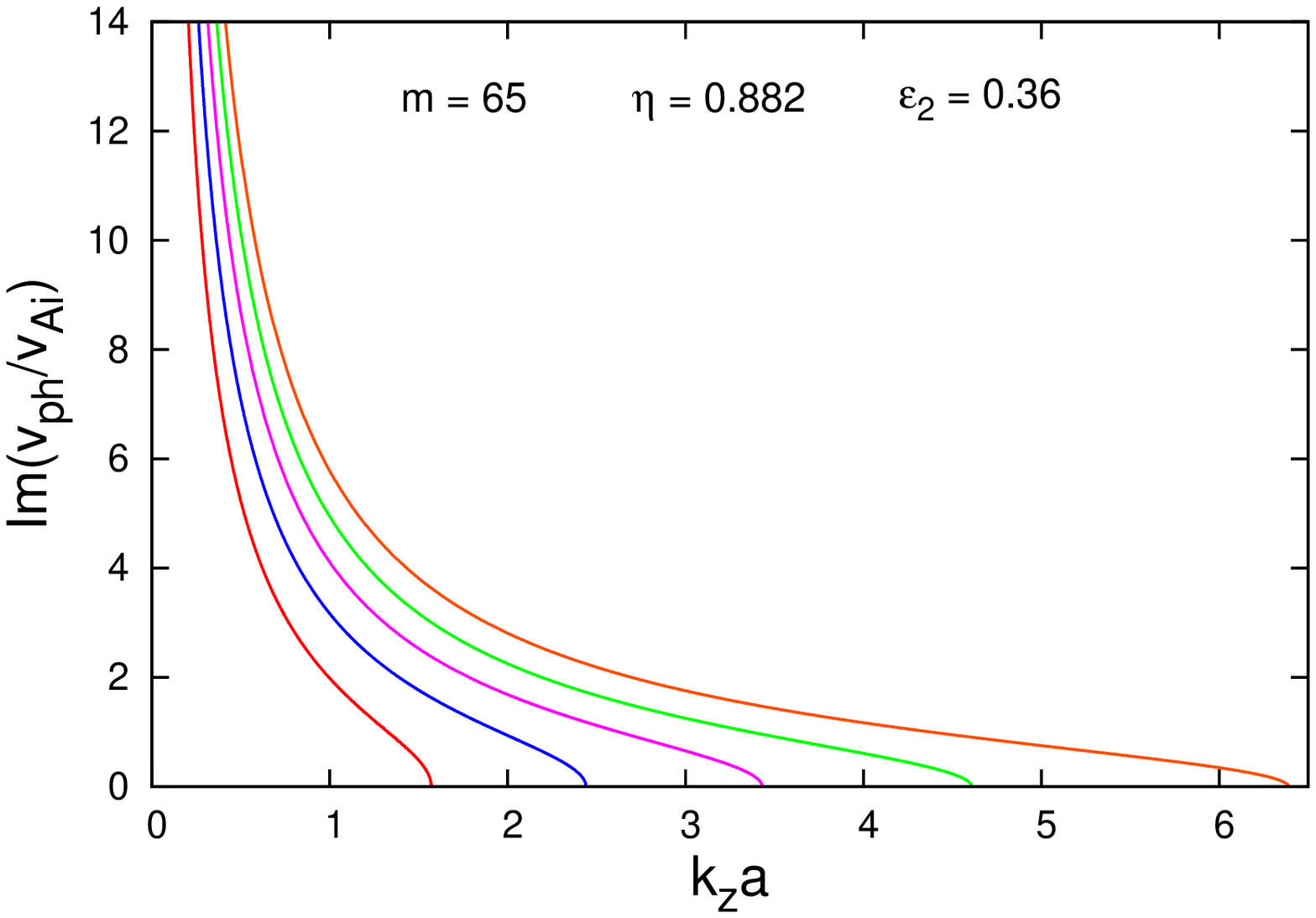}
              }
  \caption{(Left) Dispersion curves of the unstable $m = 65$ MHD mode propagating along a twisted cool jet in a coronal hole at $\eta = 0.882$, $\varepsilon_2 = 0.36$, and the following values of $\varepsilon_1$ (from right to left): $0.025$, $0.06$, $0.08$, $0.095$, and $0.10682$ (red curve in the right plot).  Alfv\'en Mach numbers for these curves are respectively $0.505$, $0.505$, $0.504$, $0.503$, and $0.5026$.  (Right) Growth rates of the unstable $m = 65$ mode for the same input parameters.  The azimuthal magnetic field that corresponds to $\varepsilon_1^\mathrm{cr} = 0.10682$ (the instability window with zero width) and stops the KHI onset is equal to $0.3$~G.}
   \label{fig:fig6}
\end{figure}

With the increase in the parameter $\varepsilon_1$, the instability region, as seen from the right panel of \textbf{Figure~\ref{fig:fig6}}, becomes narrower and at the lower limit $(k_z a)_\mathrm{lhs} = \pi/2$ with $\varepsilon_1^\mathrm{cr} = 0.10682$ and $M_\mathrm{A} = 0.5026$ its width is equal to zero.  In other words, there is no longer instability.  Therefore, the critical azimuthal magnetic field that suppresses the KHI is $B_{\mathrm{i}\phi}^\mathrm{cr} \cong 0.3$~G---obviously a relatively small value.

\subsection{Kelvin--Helmholtz Instability in a Jet Emerging from a Filament Eruption}
\label{subsec:filament}
\begin{figure}
\centering
    \includegraphics[height=.30\textheight]{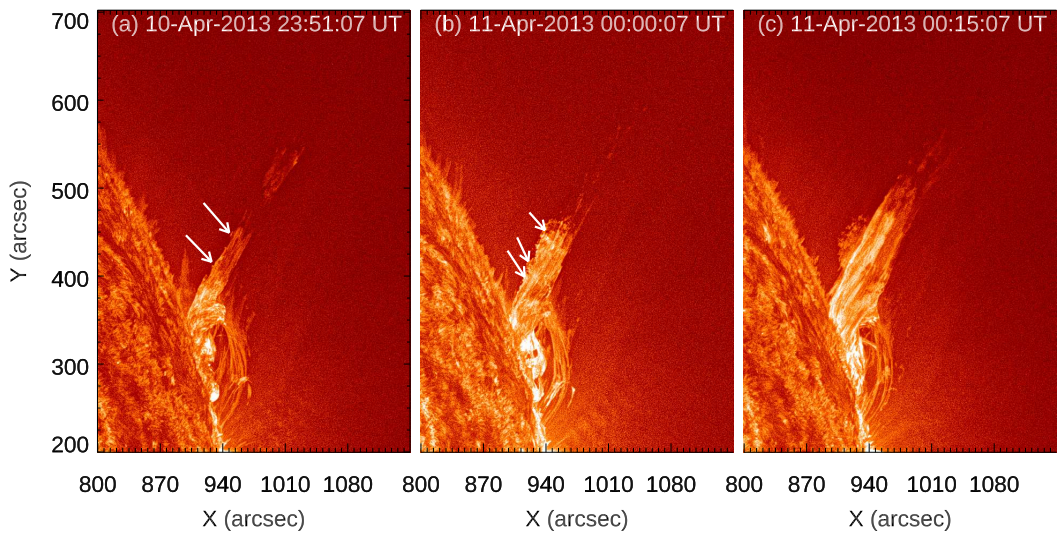}
  \caption{Evolution of the jet associated with a filament eruption observed on 2013 April 10--11 in AIA $304$~\AA{}.  The structure shown by arrows can be due to the KHI.}
   \label{fig:fig7}
\end{figure}

\citet{Filippov2015} observationally studied three jets events originated from the active region NOAA 11715 (located on the west limb) on 2013 April 10--11.  These authors claim that the jets originated from the emergence of a filament having a null-point (inverted \textsf{Y}) topology.  We have considered the second event described in that paper for a detailed study.  The jet electron number density, $n_\mathrm{i}$, and electron temperature, $T_\mathrm{i}$, both listed in \textbf{Table~1}, have been calculated by us using the techniques elaborated by \citet{Aschwanden2013}.  This technique requires the data from the six $94$, $131$, $171$, $193$, $211$, and $335$~\AA\ AIA/\emph{SDO\/} EUV channels.  In addition to the electron number densities and electron temperatures in the jet and surrounding plasma, we have also estimated the jet width as $\Delta \ell{\approx}30$~Mm, its height as $H = 180$~Mm, and have found the jet lifetime to be $30$~min.  The two important parameters, axial and azimuthal velocities, according to the observations, are $U_z = 100$ and $U_\phi = 180$~km\,s$^{-1}$, respectively.  The time evolution of the jet in AIA $304$~\AA\ is shown in \textbf{Figure~\ref{fig:fig7}} and we have observed vortex type structures in the eastern side of the jet, which are indicated by arrows.  These structures implicitly indicate for occurrence of KHI.

\begin{figure}[!ht]
   \centerline{\hspace*{0.015\textwidth}
               \includegraphics[width=0.515\textwidth,clip=]{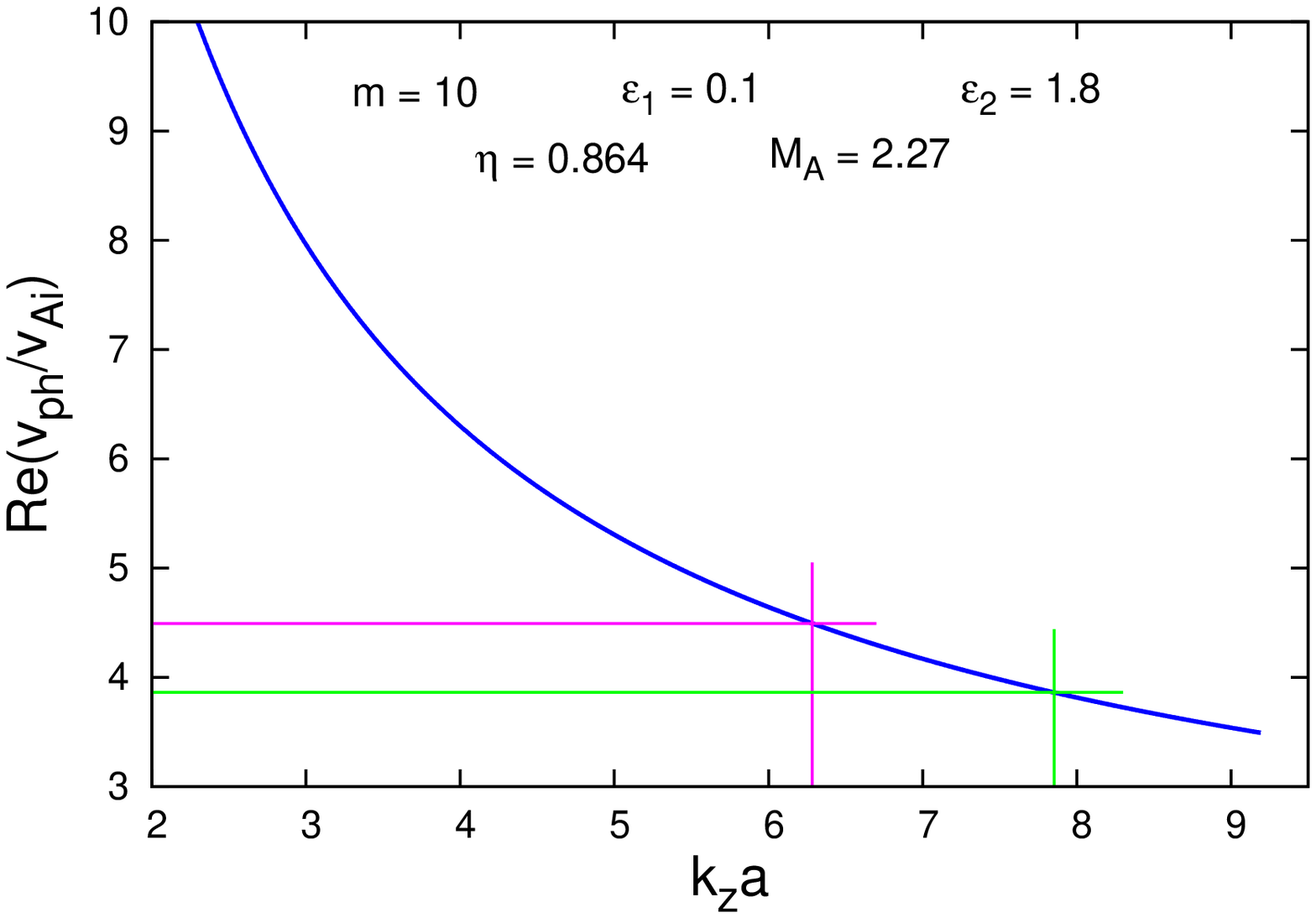}
               \hspace*{-0.03\textwidth}
               \includegraphics[width=0.515\textwidth,clip=]{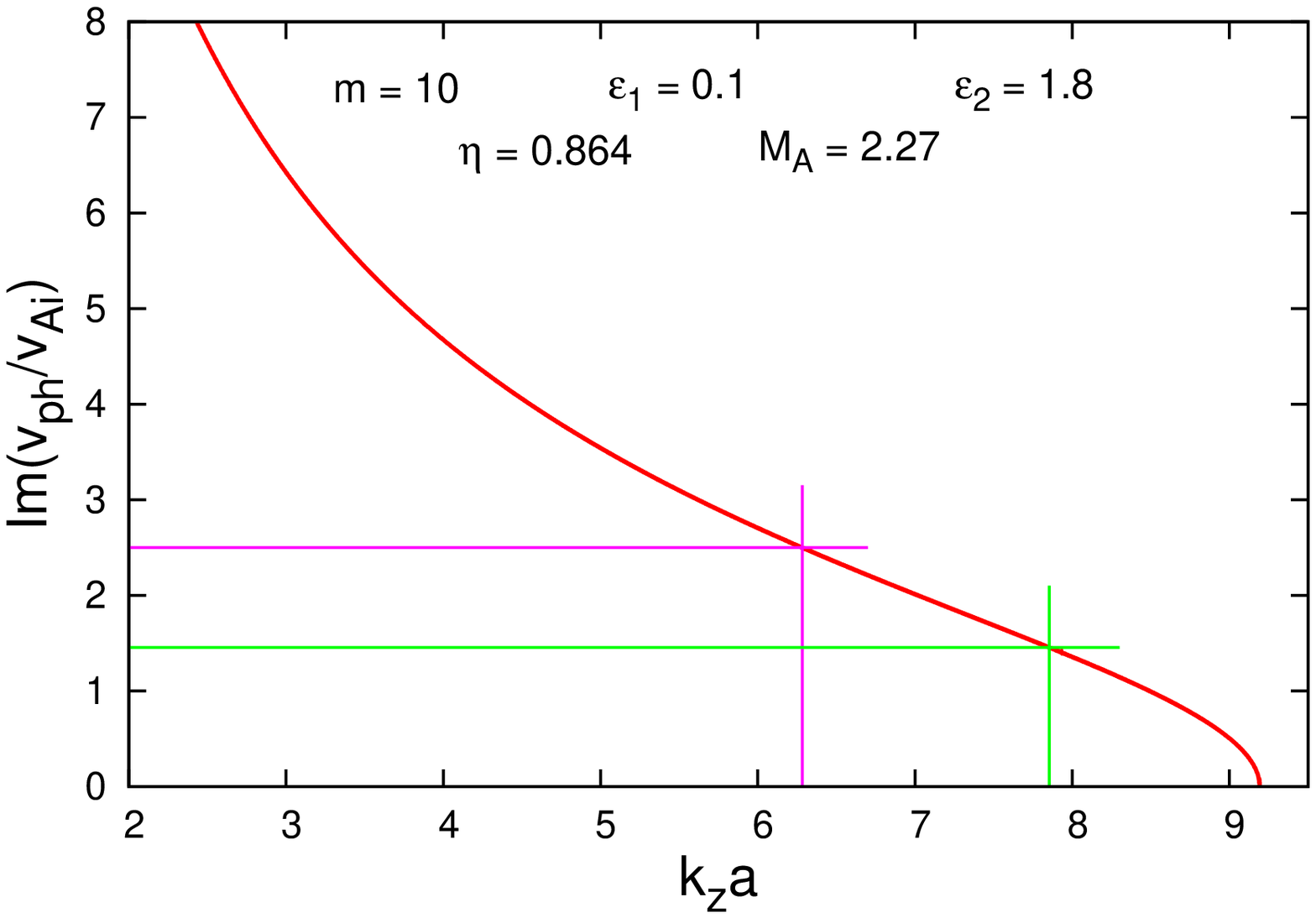}
              }
  \caption{(Left) Dispersion curve of the $m = 10$ MHD mode propagating along a twisted incompressible emerging from a filament eruption jet at $\eta = 0.864$, $b = 4.36$, $M_{\rm A} = 2.27$, $\varepsilon_1 = 0.1$, and $\varepsilon_2 = 1.8$.  (Right) Normalized growth rate curve of the $m = 10$ MHD mode computed at the same input parameters as in the left panel.  The crosses of green and purple lines yield the normalized values of the wave phase velocity and the wave growth rate at the two unstable wavelengths of $12$ and $15$~Mm, respectively.}
   \label{fig:fig8}
\end{figure}

With the typical $n$ and $T_\mathrm{e}$ (see \textbf{Table~1}), rotating velocity $U_\phi = 180$~km\,s$^{-1}$, assumed $B_\mathrm{e} = 6$~G, and $\varepsilon_1 = 0.1$,  equation~(\ref{eq:pbeq}) yields $\eta = 0.864$, $b = 4.36$, and $v_\mathrm{Ai} = 44.00$~km\,s$^{-1}$ (for comparison, the Alfv\'en speed in the environment is $v_\mathrm{Ae} = 206.3$~km\,s$^{-1}$).  We note that the choice of $\varepsilon_1$ was made taking into account the fact that the inclination of the treads of the jet in the event on 2013 April 10, detected by \emph{SDO}/AIA, yields a relationship between $B_{\mathrm{i}\phi}$ and $B_{\mathrm{i}z}$, which was evaluated as $\varepsilon_1 \approx 0.1$.  If we assume that the shortest unstable wavelength is $\lambda_\mathrm{KH} = 12$~Mm, which is located at $k_z a = 2.5\pi$ on the $k_z a$-axis, from inequality (\ref{eq:findingm}) we find that a MHD wave with $|m| = 12$ would provide an instability region, accommodating the nondimensional $k_z a = 2.5\pi$.  It turns out that a suitable mode number is $m = 10$.  The wave dispersion and growth rate diagrams are shown in \textbf{Figure~\ref{fig:fig8}}.  In that instability range one can also find the instability characteristics at $k_z a = 2\pi$, which corresponds to $\lambda_\mathrm{KH} = 15$~Mm.  The input parameters for finding the solutions to  equation (\ref{eq:dispeq}) are: $m = 10$, $\eta = 0.864$, $\varepsilon_1 = 0.1$, $\varepsilon_2 = 1.8$, $b = 4.36$, and $M_\mathrm{A} = 2.27$ (${=}100/44$) The KHI parameters at the aforementioned wavelengths are as follows:
\begin{eqnarray*}
    \gamma_\mathrm{KH} \cong 33.51 \times 10^{-3}\:\mathrm{s}^{-1}, \;\, \tau_\mathrm{KH} \cong 3.1\:\mathrm{min}, \;\, v_\mathrm{ph} \cong 170\:\mathrm{km}\,\mathrm{s}^{-1},\;\,\mbox{for} \;\,\lambda_\mathrm{KH} = 12\:\mathrm{Mm},
\end{eqnarray*}
and
\begin{eqnarray*}
    \gamma_\mathrm{KH} \cong 46.06 \times 10^{-3}\:\mathrm{s}^{-1}, \;\, \tau_\mathrm{KH} \cong 2.3\:\mathrm{min}, \;\, v_\mathrm{ph} \cong 198\:\mathrm{km}\,\mathrm{s}^{-1},\;\,\mbox{for}\;\,\lambda_\mathrm{KH} = 15\:\mathrm{Mm}.
\end{eqnarray*}
The KHI developing or growth times seem reasonable and the wave phase velocities are super-Alfv\'enic ones.

\begin{figure}[!ht]
   \centering
  \includegraphics[width=0.515\textwidth,clip=]{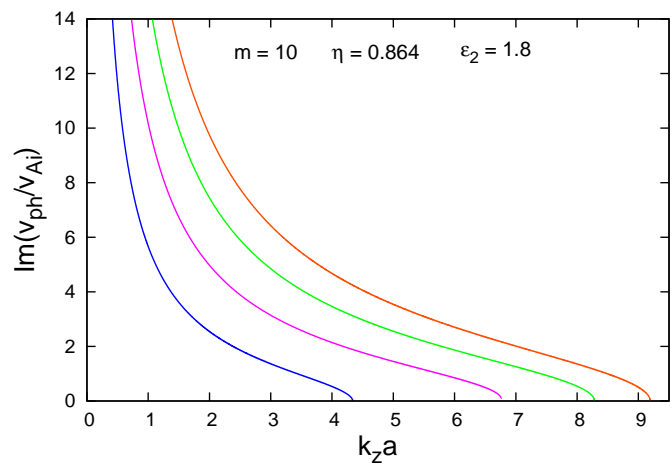}
  \caption{Growth rate curves of the unstable $m = 10$ MHD mode propagating along a twisted incompressible emerging from a filament eruption jet at $\eta = 0.864$, $\varepsilon_2 = 1.8$, and the following values of $\varepsilon_1$ (from right to left): $0.1$ (orange), $0.6$ (green), $0.8$ (purple), and $0.9$ (blue)}.
   \label{fig:fig9}
\end{figure}

It is intriguing to see how the width of the instability range will shorten as the magnetic field twist $\varepsilon_1$ is increased.  Our numerical computations indicate that for a noticeable contraction of the instability window one should change the magnitude of $\varepsilon_1$ with relatively large steps.  The results of such computations are illustrated in \textbf{Figure~\ref{fig:fig9}}.  It is necessary to underline that at values of $\varepsilon_1$ close to $1$, (i) $\beta_\mathrm{i}$ becomes ${<}1$ and the jet has to be treated as a cool medium, which implies a new wave dispersion relation and probably a higher wave mode number, $m$; (ii) one cannot use $\varepsilon_1 > 1$, because in that case the instability is of another kind, namely kink instability \citep{Lundquist1951,Hood1979,Zaqarashvili2014}.  At this `pathological' case, one cannot reach the lower limit of the instability range, $(k_z a)_\mathrm{lhs} = 0.524$, and consequently we are unable to evaluate that azimuthal magnetic field, $B_{\phi}^\mathrm{cr}$, which will stop the KHI onset!

\subsection{Kelvin--Helmholtz Instability in a Spinning Macrospicule}
\label{subsec:macrospicule}
As we have mentioned in Section~1, \citet{Pike1998} did a statistical study of the dynamics of solar transition region features, like macrospicules.  These features were observed on the solar disk and also on the solar limb by using data from the \emph{Coronal Diagnostic Spectrometer\/} (CDS) onboard \emph{SOHO}. In addition, in their article, \citet{Pike1998} discussed the unique CDS observations of a macrospicule first reported by \citet{Pike1997} along with their own (Pike and Mason) observations from the \emph{Normal Incidence Spectrometer\/} (NIS).  This spectrometer covers the wavelength range from $307$ to $379$~\AA{} and that from $513$ to $633$~\AA{} using a microchannel plate and CCD combination detector.  The details of macrospicule events observed near the limb are given in Table I in \citealp{Pike1998}, while those of macrospicule events observed on the disk are presented in Table II.  The main finding in the study of \citet{Pike1998} was the rotation in these features based on the red and blue shifted emission on either side of the macrospicule axes.  According to the authors, the detected rotation assuredly plays an important role in the dynamics of the transition region.  Using the basic observational parameters obtained by \citet{Pike1998}, \citet{Zhelyazkov2019} examined the conditions for KHI rising in the macrospicule.  Let us discuss that study as follows.

Our (\citealp{Zhelyazkov2019}) choice for modeling namely the macrospicule observed on 1997 March 8 at 00:02~\textsc{ut} (see Table II in \citealp{Pike1998}) was made taking into account the fact that this macrospicule possesses the basic characteristics of the observed over the years tornado-like jets---the axial velocity of the jet was $U_z =75$~km\,s$^{-1}$, while its rotating speed, we evaluate to be $U_\phi = 40$~km\,s$^{-1}$.  For the other characteristics of the macrospicule such as lifetime, maximum width, average flow velocity, and maximum length or height, we used some average values obtained from a huge number of observations and specified in \citealp{Kiss2017} as $16.75 \pm 4.5$~min, $6.1 \pm 4$~Mm, $73.14 \pm 25.92$~km\,s$^{-1}$, and $28.05 \pm 7.67$~Mm, respectively.  For our study here, we take the macrospicule width to be $\Delta \ell = 6$~Mm, its height $H = 28$~Mm, and lifetime of the order on $15$~min.  The basic macrospicule physical parameters (see \textbf{Table~1}) with $U_\phi = 40$~km\,s$^{-1}$, $\varepsilon_1 = 0.005$, and $B_\mathrm{e} = 5$~G yield (using equation (\ref{eq:pbeq})) $\eta = 0.1$, Alfv\'en speed $v_\mathrm{Ai} = 60.6$~km\,s$^{-1}$ (while in the surrounding plasma we have $v_\mathrm{Ae} \cong 345$~km\,s$^{-1}$), and $b = 1.798$.  The excited MHD mode, whose instability window would contain a $\lambda_\mathrm{KH} = 3$~Mm, is $|m| = 52$.  Performing the numerical computations with the input parameters: $m = 52$, $\eta = 0.1$, $\varepsilon_1 = 0.005$, $\varepsilon_2 = 0.53$ (${=}40/75$), $b = 1.798$, and $M_\mathrm{A} = 1.24$ (${=}75/60.6$), we get plots very similar to those pictured in \textbf{Figure~\ref{fig:fig2}}, which (the plots) allow us to find the KHI characteristics for the two wavelengths of $3$ and $5$~Mm, exactly
\begin{eqnarray*}
    \gamma_\mathrm{KH} \cong 48.38 \times 10^{-3}\:\mathrm{s}^{-1}, \;\, \tau_\mathrm{KH} \cong 2.2\:\mathrm{min}, \;\, v_\mathrm{ph} \cong 361\:\mathrm{km}\,\mathrm{s}^{-1},\;\,\mbox{for} \;\,\lambda_\mathrm{KH} = 3\:\mathrm{Mm},
\end{eqnarray*}
and
\begin{eqnarray*}
    \gamma_\mathrm{KH} \cong 184.8 \times 10^{-3}\:\mathrm{s}^{-1}, \;\, \tau_\mathrm{KH} \cong 0.57\:\mathrm{min}, \;\, v_\mathrm{ph} \cong 556\:\mathrm{km}\,\mathrm{s}^{-1},\;\,\mbox{for}\;\,\lambda_\mathrm{KH} = 5\:\mathrm{Mm}.
\end{eqnarray*}
One observes that at both unstable wavelengths the corresponding phase velocities are supper-Alfv\'enic.  Moreover, the two growth times of $2.2$ and ${\sim}0.6$~min seem reasonable bearing in mind the fact that the macrospicule lifetime is about $15$ minutes, which implies that the KHI at the selected wavelengths is rather fast.  The $B_{\mathrm{i}\phi}^\mathrm{cr}$ that suppresses the KHI onset, equals $0.57$~G and was calculated with $\varepsilon_1^\mathrm{cr} = 0.202085$ and $M_\mathrm{A} = 1.2119$.  Our study \citep{Zhelyazkov2019} shows that a decrease in the background magnetic field to $B_\mathrm{e} = 4.8$~G would require the excitation of MHD wave with mode number $m = 48$, at which the KHI characteristics at the wavelengths of $3$ and $5$~Mm are very close to those obtained with $m = 52$.

\section{Summary and Outlook}
\label{sec:outlook}
In this article, we have studied the emerging of KHI in four different spinning solar jets (standard and blowout coronal hole jets, jet emerging from a filament eruption, and rotating macrospicule) due to the excitation of high-mode ($m \geqslant 2$) MHD waves traveling along the jets.  First and foremost, we model each jet as a vertically moving with velocity $\boldsymbol{U}$ cylindrical twisted magnetic flux tube with radius $a$.  There are four basic steps in the modeling as follows:
\begin{itemize}
\item \emph{Topology of jet--environment magnetic and velocity fields\/} For simplicity, we assume that the plasma densities of the jet and its environment, $\rho_\mathrm{i}$ and $\rho_\mathrm{e}$, respectively, are homogeneous.  Generally they are different and the density contrast is characterized by the ratio $\rho_\mathrm{e}/\rho_\mathrm{i} = \eta$.  The twisted internal magnetic and velocity fields are supposed to be uniform, that is, represented in cylindrical coordinates, $(r, \phi, z)$, by the vectors $\boldsymbol{B}_\mathrm{i} = (0, B_{\mathrm{i}\phi}(r), B_{\mathrm{i}z})$ and $\boldsymbol{U} = (0, U_\phi(r), U_z)$, where their azimuthal components are considered to be linear functions of the radial position $r$, \emph{viz}.\ $B_{\mathrm{i}\phi}(r) = Ar$ and $U_\phi(r) = \Omega r$, where $A$ and $\Omega$ (the azimuthal jet velocity) are constants.  We note that $B_{\mathrm{i}z}$ and $U_z$ are also constants.  It is convenient the twists of the magnetic field and the flow velocity of the jet to be characterized by the two numbers $\varepsilon_1 = B_{\mathrm{i}\phi}(a)/B_{\mathrm{i}z} \equiv Aa/B_{\mathrm{i}z}$ and $\varepsilon_2 = U_\phi(a)/U_z \equiv \Omega a/U_z$, respectively.  Note that $\Omega a$ is the jet rotational speed $U_z$.  The surrounding coronal or chromospheric plasma is assumed to be immobile and embedded in a constant magnetic field $\boldsymbol{B}_\mathrm{e} = (0,0,B_\mathrm{e})$.  In our study, the density contrast, $\eta$, varies from $0.1$ to $0.9$, the magnetic field twist, $\varepsilon_1$, can have a wide range of magnitudes from $0.005$ to $0.95$ (it has to be less than $1$ in order to avoid the rising of the kink instability), while the velocity twist parameter, $\varepsilon_2$, is fixed by the observationally measured rotational and axial speeds.

\item \emph{Listing of the basic physical parameters and determination of plasmas betas\/}  In general, at a fixed density contrast, the plasma beta is controlled by the magnetic field (inside or outside the magnetic flux tube) and the electron temperatures of the jet and surrounding plasma.  The values of these physical parameters should satisfy the total pressure balance equation~(\ref{eq:pbeq}) at all levels (equilibrium and perturbational).  Our practice is to fix $B_\mathrm{e}$, and  using equation~(\ref{eq:pbeq}) to determine the internal Alfv\'en speed defined as $v_\mathrm{Ai} = B_{\mathrm{i}z}/\sqrt{\mu \rho_\mathrm{i}}$.  It is worth underlying that the usage of equation~(\ref{eq:pbeq}) requires the specification of $\varepsilon_1$.  In our four cases, for finding the KHI characteristics, we took $\varepsilon_1$ to be equal to $0.005$, $0.025$, or $0.1$.  The electron temperatures in the jets are from $500\,000$~K for the macrospicule to $2.0$~MK in the jet emerging from a filament eruption.  The electron temperatures of surrounding plasmas are $2.14$~MK in the active solar region \citep{Filippov2015}, $2.0$~MK in the blowout coronal hole jet \citep{Young2014a}, and $1.0$~MK in the environments of the standard coronal hole jet \citep{Chen2012} and the macrospicule.  With background magnetic fields of $3$ to $6$~G, rotating velocities of $40$ to $180$~km\,s$^{-1}$, and $\varepsilon_1 = 0.005$, the total pressure balance equation (\ref{eq:pbeq}) yields plasma betas of the first, third, and forth jets greater than $1$ and those of the environments and the internal medium of the second jet much less than $1$ (see \textbf{Table~1}).  With these plasma beta values one can consider the media of the standard coronal hole jet, the rotating jet emerging from a filament eruption, and microspicule as nearly incompressible plasmas, while the internal medium of the blowout coronal hole jet and the surrounding magnetized plasma in the four cases can be treated as a cool medium \citep{Zank1993}.

\item \emph{Solving of the wave dispersion relation and finding the KHI characteristics\/}  For finding the solutions to the MHD wave dispersion equations~(\ref{eq:dispeq}) or (\ref{eq:dispeqc}), which are a slight modification of the `basic' dispersion relation derived in \citealp{Zaqarashvili2015}, it is necessary to specify the following input data: the wave mode number, $m$, the density contrast, $\eta$, the two twist parameters $\varepsilon_1$ and $\varepsilon_2$, the magnetic fields ratio, $b$ (obtainable from equation~(\ref{eq:pbeq})), and the Alfv\'en Mach number $M_\mathrm{A} = U_z/v_\mathrm{Ai}$.  The roots of the dispersion equations~(\ref{eq:dispeq}) or (\ref{eq:dispeqc}) are the normalized wave phase velocity and instability growth rate as functions of the nondimensional wavenumber $k_z a$.  From plots which graphically represent the found solutions, one obtains at the anticipated wavelengths (given by their $k_z a$-values on the horizontal $k_z a$-axes) the corresponding $\mathrm{Re}(v_\mathrm{ph}/v_\mathrm{Ai}$) and $\mathrm{Im}(v_\mathrm{ph}/v_\mathrm{Ai}$) values.  From them, one can find in absolute units the KHI growth rate, $\gamma_\mathrm{KH}$, the instability developing or growth time, $\tau_\mathrm{KH} = 2\pi/\gamma_\mathrm{KH}$, and the corresponding wave phase velocity, $v_\mathrm{ph}$.  The MHD wave mode numbers at which we were able to calculate the instability characteristics at wavelengths comparable to the radius or width of the jet are between $10$ and $65$ and the KHI growth times at those wavelengths are of the order on a few minutes, generally in good agreement with the observations.  It is curious to note that in searching KHI growth times of the order on few seconds, when studying the dynamics and stability of small-scale rapid redshifted and blueshifted excursions, appearing as high-speed jets in the wings of the H$\alpha$ line, \citet{Kuridze2016} had to assume the excitation of MHD waves with mode numbers up to $100$.  A typical property of the instability developing times, owing to the shape of the plotted dispersion curves, is that with increasing the examined wavelength the growth times become shorter---for instance, at $\lambda_\mathrm{KH} = 10$~Mm the KHI developing time in the coronal hole jet \citep{Chen2012} is around $4.5$~min, while at $\lambda_\mathrm{KH} = 12$~Mm it is equal to ${\cong}2.1$~min.  A change in  $B_\mathrm{e}$ can influence the MHD mode number $m$, which would yield an instability region similar or identical to that seen in \textbf{Figure~\ref{fig:fig2}}.  It is necessary to mention that the width of the instability range except by changing the MHD wave mode number, $m$, can be regulated by increasing or diminishing the parameter $\varepsilon_1$.

\item \emph{Finding the critical azimuthal magnetic field which suppresses the emergence of KHI\/}  It was numerically established, that any increase in $\varepsilon_1$ yields to the shortening of the instability range.  This observation implies that there should exist some critical $\varepsilon_1$, at which the upper limit of the instability range coincides with the lower one---in that case the width of the instability window is zero, which means that there is no longer any instability.  With such an $\varepsilon_1^\mathrm{cr}$, one can calculate the $B_{\mathrm{i}\phi}^\mathrm{cr}$, which stops the KHI appearance.  For the rotating blowout coronal hole jet this $B_{\mathrm{i}\phi}^\mathrm{cr}$ is relatively small---it is equal to ${\approx}0.3$~G, while for the standard coronal hole jet it is $1.4$~G.  It is worth noticing that due to the specific parameters of the jet emerging from a filament eruption, we were unable to find a $B_{\mathrm{i}\phi}^\mathrm{cr}$, which would stop the KHI onset because with values of $\varepsilon_1$ close to $1$ our dispersion relation becomes inappropriate (the internal medium being nearly incompressible becomes a cool one) and we cannot calculate that $\varepsilon_1$ at which the lower limit, $(k_z a)_\mathrm{lhs} = 0.524$, can be reached (see \textbf{Figure~\ref{fig:fig9}}).
\end{itemize}

In this article, we also corrected the total pressure balance equation used in \citealp{Zhelyazkov2018a} and \citealp{Zhelyazkov2018b}, which turns out to be erroneous.  The true total balance equation is given by equation~(\ref{eq:pbeq}).  In fact, the corrected pressure balance equation, used here, changes the mode numbers at which the KHI occurs, namely from $m = 12$ to $m = 11$ for the coronal hole jet and from $m = 18$ to $m = 10$ for the rotating jet emerging from a filament eruption.  The computed KHI developing or growth times in the aforementioned articles, nonetheless, are not changed noticeably---they are of the same order with these computed in this paper.  In addition, there is another improvement issue, scilicet when studying how the increasing $\varepsilon_1$ shortens the instability region, one has to apply equation~(\ref{eq:pbeq}) for obtaining the appropriate values of $b = B_\mathrm{e}/B_{\mathrm{i}z}$ and $v_\mathrm{Ai} = B_{\mathrm{i}z}/\sqrt{\mu \rho_\mathrm{i}}$, and subsequently $M_\mathrm{A} = U_z/v_\mathrm{Ai}$.  Thus, one can claim that equation~(\ref{eq:pbeq}) plays an important role in the numerical studies of the KHI in various solar jets.

Our approach in investigating the KHI in rotating twisted solar jets can be improved in the following directions: (i) to assume some radial profile of the plasma density of the jet, which immediately will require additional study on the occurrence of continuous spectra and resonant wave absorption \citep{Goedbloed2004}, alongside to see to what extent these phenomena will influence the instability growth times; (ii) to investigate the impact of the nonlinear azimuthal magnetic and velocity fields radial profiles on the emergence of KHI; and (iii) to derive a MHD wave dispersion relation without any simplifications like considering the jet and its environment as incompressible or cool plasmas---this will show how the compressibility will change the picture.  We should also not forget that the nonlinearity, as \citep{Miura1982,Miura1984} claim, can lead to the saturation of the KHI growth, and to formation of nonlinear waves.  Nevertheless, even in its relatively simple form, our way of investigating the conditions under which the KHI develops is flexible enough to explore that event in any rotating solar jet in case that the basic physical and geometry parameters of the jet are provided by observations.

\section*{Conflict of Interest Statement}

The authors declare that the research was conducted in the absence of any commercial or financial relationships that could be construed as a potential conflict of interest.

\section*{Author Contributions}

IZh wrote the substantial parts of the manuscript.  RJ wrote the Introduction section and prepared three figures associated with the observations.  RC contributed by writing the parts devoted to observations, as well as in the careful proofreading of the text.

\section*{Funding}
The work of IZh and RC was supported by the Bulgarian Science Fund contract DNTS/INDIA 01/7.  RJ was funded by the Department of Science and Technology, New Delhi, India as an INSPIRE fellow.  RC was also supported from the SERB-DST project no.\ SERB/F/7455/2017-17.

\section*{Acknowledgments}
We thank Professor Teimuraz V.\ Zaqarashvili for useful discussions and are deeply grateful to the reviewers for their constructive comments.

\bibliographystyle{frontiersinSCNS_ENG_HUMS} 

\newpage

\section*{Figure captions}


\clearpage
\mbox{~}
\section*{Tables}

\begin{table}
\label{tab:table1}
\caption{\textsf{Jets physical parameters derived from observational data}.}
\label{tab:parameter}
\vspace*{2mm}
\renewcommand*{\arraystretch}{1.4}
\centering

\begin{tabular}{lccccccc}
\hline
\textbf{\textsf{Kind of}} & \textsf{\textbf{\textit{B}}$_{\textsf{\textbf{e}}}$} & \textsf{\textbf{\textit{n}}$_{\textsf{\textbf{e}}}$} & \textsf{\textbf{\textit{n}}$_{\textsf{\textbf{i}}}$} & \textsf{\textbf{\textit{T}}$_{\textsf{\textbf{e}}}$} & \textsf{\textbf{\textit{T}}$_{\textsf{\textbf{i}}}$} &
\textsf{\textbf{$\tens{\beta}_{\textsf{\textbf{e}}}$}} &
\textsf{\bm{$\beta$}$_{\textsf{\textbf{i}}}$} \\
  \textbf{\textsf{jet}}  & \textbf{\textsf{(G)}} & \textbf{\textsf{(\textsf{\textbf{$\times$}}\textsf{\textbf{10}}$^\textsf{\textbf{9}}$ cm$^{\textbf{\textsf{--3}}}$)}} & \textbf{\textsf{(\textsf{\textbf{$\times$}}\textsf{\textbf{10}}$^\textsf{\textbf{9}}$ cm$^{\textbf{\textsf{--3}}}$)}} & \textbf{\textsf{(MK)}} & \textbf{\textsf{(MK)}} &  & \\
\hline
\textsf{Standard coronal hole}  & \textsf{3.0} & \textsf{0.90}  & \textsf{1.00}  & \textsf{1.00}  & \textsf{1.6} & \textsf{0.348} & \textsf{2.079} \\
\textsf{Blowout coronal hole}  & \textsf{3.0} & \textsf{0.15}  & \textsf{0.17}  & \textsf{2.00}  & \textsf{1.7} & \textsf{0.116} & \textsf{0.115} \\
\textsf{Filament eruption} & \textsf{6.0} & \textsf{4.02} & \textsf{4.65} & \textsf{2.14} & \textsf{2.0} & \textsf{0.831} & \textsf{17.24} \\
\textsf{Macrospicule} & \textsf{5.0} & \textsf{0.10}  & \textsf{1.00}  & \textsf{1.00}  & \textsf{0.5} & \textsf{0.139} & \textsf{2.248} \\
\hline
\end{tabular}
\end{table}
%


\end{document}